\documentclass[natbib,twocolumn]{svjour3}         
\smartqed  
\usepackage{graphicx}
\usepackage{mathptmx}      
\newcommand{\rp}{r_{\mathrm p}}
\newcommand{\mpl}{M_{\mathrm p}}

\newcommand{\Sp}{\Sigma_{\mathrm p}}
\newcommand{\Rp}{\rho_{\mathrm p}}
\newcommand{\op}{\Omega_\mathrm{p}}

\newcommand{\me}{\mathrm{M_\oplus}}
\newcommand{\xs}{x_\mathrm{s}}

\journalname{Earth, Moon and Planets}
\begin{document}

\title{Numerical simulations of disc-planet interactions
}


\author{Richard P. Nelson \and
        Sijme-Jan Paardekooper
}


\institute{R.P. Nelson \at Astronomy Unit, Queen Mary University of London,
                           Mile End Rd, London, E1 4NS, U.K. \\
                        \email{R.P.Nelson@qmul.ac.uk}
      \and S.-J. Paardekooper \at DAMTP, Wilberforce Road, 
                                  Cambridge CB3 0WA, U.K.  \\
              \email{S.Paardekooper@damtp.cam.ac.uk} 
}

\date{Received: date / Accepted: date}

\maketitle

\begin{abstract}
The gravitational interaction between a protoplanetary disc and planetary sized
bodies that form within it leads to the exchange of angular momentum,
resulting in migration of the planets and possible gap formation in the disc
for more massive planets. In this article, we review the basic theory of
disc-planet interactions, and discuss the results of 
recent numerical simulations 
of planets embedded in protoplanetary discs. We consider the migration of low
mass planets and recent developments in our understanding of so-called type I
migration when a fuller treatment of the disc thermodynamics is included. 
We discuss the runaway migration of intermediate mass planets
(so-called type III migration), and the migration of giant planets
(type II migration) and the associated gap formation in this disc.
The availability
of high performance computing facilities has enabled global 
simulations of magnetised,
turbulent discs to be computed, and we discuss recent results for both low
and high mass planets embedded in such discs. 

\end{abstract}

\section{Introduction}
\label{intro}
Planets are believed to form in the circumstellar discs of gas and dust
that are observed around young stars \cite{becksar96}. Within these
discs, sub-micron sized interstellar dust grains collide and stick,
slowly growing to form bodies comparable in mass to the Earth ($\me$)
{\it via} a multi-stage process which involves bodies of 
increasing size colliding and sticking. These planetary bodies which
form with masses up to $\simeq 10$ $\me$ can accrete gas envelopes,
leading to the formation of a gas giant planet \cite{pollack96},
a process which, models predict, requires a few million years.
The precursor $\simeq 10$ $\me$ bodies which accrete gas
are usual referred to as planetary cores, and their 
formation is most efficient at locations in the disc where
the density of solids is highest. Thus gas giant planets are thought to form 
in the cool disc regions beyond the `snowline', where volatiles can condense
into ice grains. In most disc models, and apparently in our own
early Solar System, the snowline is located at a radius $\sim 3$ AU
from the central star.
The discovery of the first extrasolar planet in 1995 \cite{mayor}, a
gas giant planet orbiting at $0.05$ Astronomical Units (AU) from the star 51 Pegasi
therefore came as a big surprise, and posed a problem for planet
formation theory. 

The solution to this problem is thought to be that planets such as 
51 Peg b probably did form several AU away from the star,
but migrated inward afterwards, while still embedded in the
circumstellar disc. The large number of extrasolar planets now known
(more than 350 at the time of writing), with broad variation
in their masses and orbital configurations,
indicates that migration is a common phenomenon during planet formation.

Prior to the discovery of extrasolar planets, it was already known
that embedded bodies experience a force from
the surrounding gas disc \cite{gt79, gt80, linpap1979, linpap1986} 
leading to orbital
migration. This can be appreciated in a qualitative way from
Fig. \ref{fig2}, where we show the surface density perturbation in a
disc by a low-mass planet. The planet launches two tidal waves that
are sheared into spiral waves by differential rotation in the
disc. The gravitational pull from these waves leads to orbital
migration of the planet.

Disc-planet interaction has become the focus of intense
research since 1995, and three types of migration can be distinguished: Type I
migration for low-mass planets (comparable to the Earth), which do not
strongly perturb the surrounding disc, Type II migration for high-mass
planets (comparable to Jupiter), which open deep gaps around their
orbit, and Type III migration for intermediate-mass planets
(comparable to Saturn), embedded in massive discs \cite{maspap}. 

A significant fraction of the research in this area has utilised
numerical simulations, and a number of insights have been gained
from the results of such simulations which have subsequently
been intepreted within the fraemwork of an analytic model.
The increasing availability of parallel computing facilities,
combined with the development of simulation codes, has allowed
increasingly complex simulations to be performed, including 
physics such as MHD turbulence and radiation transport.
In this article we will review the
basic theory of disc-planet interactions, highlight some recent
developments, and summarise the state of the field as it
currently stands.

\section{Type I migration}
\label{typeI}
Type I migration is thought to apply to low-mass planets
(approximately 1-20 $\me$). It is important to understand this regime,
because not only does it apply to terrestrial planets, also the solid
cores of gas giant planets have been subject to Type I migration
before they obtained their gaseous envelope. This stage in giant
planet formation requires more than $10^6$ years \cite{pollack96}. In this
section, we will review the basic theory of Type I migration, 
focusing on laminar (non-turbulent) discs.

Type I migration has long been regarded as the simplest migration
mechanism to understand. Since low-mass planets induce only small
perturbations in the disc, a linear analysis should provide accurate
estimates of the migration rate \cite{tanaka}. These can later be
verified by numerical hydrodynamical simulations
\cite[e.g.][]{nelson04}. 

\subsection{Linear analysis}
It is convenient to express the perturbing potential due to the planet
acting on the disc $\Phi_\mathrm{p}$ as a Fourier series
\begin{equation}
\Phi_\mathrm{p}(r,\varphi,t) = -\frac{G\mpl}{|\vec r -\vec \rp|}=\sum_{m=0}^{\infty}
\Phi_m(r)\cos(m\varphi-m\op t), 
\end{equation}
where $\varphi$ is the azimuthal angle and $\op$ is the orbital
frequency of the planet, and to evaluate the response of the disc due to
each component separately. For simplicity, we will consider only
two-dimensional, isothermal discs in this section. The vertically
integrated pressure is then given by $P=c^2\Sigma$, where $c$ is the
sound speed and $\Sigma$ is the surface density of the disc
defined by $\Sigma=\int^{\infty}_{-\infty} \rho \, dz$. In
vertical hydrostatic equilibrium, the pressure scale height of the
disc equals $H=c/\Omega$, where $\Omega$ is the angular velocity of the
disc. A typical disc has $h\equiv H/r \approx 0.05$. The unperturbed
surface density follows a power law in radius, $\Sigma \propto
r^{-\alpha}$. 

When working with discs, angular momentum is a key quantity,
dominating the dynamics. The angular momentum of the planet is
$J_\mathrm{p}=\mpl\rp^2\op \propto \sqrt{\rp}$ for a circular
orbit. Therefore, changes in angular momentum are directly related to
a change in orbital radius, which makes the torque, the change in
angular momentum, an important quantity to measure.

Writing the perturbed surface density as 
\begin{equation}
\Sigma'(r,\phi,t) = \Sigma'_m(r)\exp(im\varphi-im\op t),
\end{equation}
one can numerically solve the linearised fluid equations to
obtain the disc response. The torque on the planet,
\begin{equation}
\Gamma=\int_\mathrm{Disc} \Sigma \vec{r} \times \nabla \Phi_\mathrm{p}
dS,
\end{equation}   
is directly related to the imaginary part of the surface density
perturbation:
\begin{equation}
\frac{d\Gamma_m}{dr}=-\pi m \Phi_m {\cal I}m(\Sigma'_m),
\end{equation}
and the torque $\Gamma_m$ can be found by integrating over the
whole disc. The total torque is equal to the sum over all $\Gamma_m$.  

Neglecting pressure effects, the disc can exchange angular momentum
with the planet at a Lindblad resonance, which is
located at a position in the disc where $m(\Omega-\op)=\pm
\kappa$, or at a corotation resonance, where $\Omega=\op$. Here,
$\kappa$ is the epicyclic frequency, equal to $\Omega$ in a Keplerian
disc. At a Lindblad resonance, a pressure wave is excited that
propagates away from the planet, which cause an oscillatory
disc response in the solutions to linearised equations described above.
The waves transport angular momentum, resulting in a torque on the
planet. The outer wave removes angular momentum from the planet, while
the inner wave gives angular momentum to the planet. The torque
therefore depends on the difference in the strengths of the two waves. 

There are several effects that conspire to make the outer wave
stronger \cite{ward97}, the most important one being that outer
Lindblad resonances are located closer to the planet than inner
Lindblad resonances. Therefore, for all reasonable surface density
profiles the total Lindblad torque on the planet is negative,
resulting in inward migration. The most recent linear calculations
for a 2D disc \cite{tanaka} resulted in a Lindblad torque
\begin{equation}
\label{eqLind}
\Gamma_\mathrm{L,lin} = -\left(3.2+1.468\alpha\right)
\frac{q^2}{h^2}\Sp\rp^4 \op^2,
\end{equation}  
where $q$ denotes the mass of the planet in units of the central mass,
$\rp$ denotes the orbital radius of the planet and $\Sp$ is the
surface density at the location of the planet. Note that $\alpha$ is
usually thought to be positive, resulting in a negative torque on the
planet.    
  
\begin{figure}
\includegraphics[width=0.8\columnwidth]{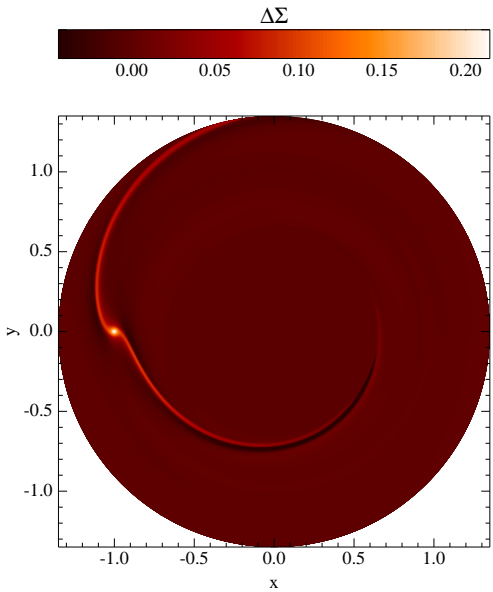}
\caption{Surface density perturbation for a 4 $\me$ planet (located at
  $(x,y)=(-1,0)$) embedded in a disc with $h=0.05$, showing the
  prominent spiral wakes associated with Lindblad torques.}  
\label{fig2}
\end{figure}

In figure \ref{fig2} we show the surface density perturbation due to a
low-mass planet from a numerical hydrodynamical calculation. Clearly
visible are the two spiral waves launched by the planet that result in
the above Lindblad torque. 

As mentioned above, angular momentum exchange also takes place at a
corotation resonance. Since the location of this resonance is very
close to the orbit of the planet, the resulting corotation torque is
potentially very strong. Since the relative velocity of the gas and
the planet is subsonic near the corotation resonance, the planet is
unable to excite waves at this location \cite{goodman}. 
Semi-analytical calculations
\cite{tanaka} result in a linear corotation torque  
\begin{equation}
\label{eqCor}
\Gamma_\mathrm{c,lin} = 1.36\left(\frac{3}{2}-\alpha\right)
\frac{q^2}{h^2}\Sp\rp^4 \op^2,
\end{equation}  
and therefore a total linear torque
\begin{equation}
\Gamma_\mathrm{lin}=\Gamma_\mathrm{L,lin} +
\Gamma_\mathrm{c,lin}=-\left(1.16+2.828\alpha\right)
\frac{q^2}{h^2}\Sp\rp^4 \op^2.
\end{equation} 
Again, since $\alpha$ is usually thought to be positive, we find that
low-mass planets will migrate inward. In equation \ref{eqCor}, the
quantity in brackets is the power law index of the specific
vorticity, or vortensity $(\nabla\times\vec{v})/\Sigma$, where
$\vec{v}$ is the velocity in the disc. The
corresponding migration rate is (assuming a circular orbit):
\begin{equation}
\dot \rp = \frac{2\Gamma_\mathrm{lin}}{\rp\op\mpl}=
-\left(1.16+2.828\alpha\right) \frac{q}{h^2}\frac{\Sp\rp^2}{M_*}\rp
\op.  
\end{equation}
 
A major problem with the resulting inward migration rates is that they
are much too high: Type I migration would quickly take all low-mass
planets, including the cores of gas giant planets, very close to or
even into the central star \cite{ward97}. Models of planet formation
including Type I migration \cite{idalin} have great difficulty
explaining the observed distribution of planets. 

Several mechanisms have been proposed to slow down Type I
migration. Within linear theory, it is first of all important to note
that the two-dimensional approximation is not valid once the
gravitational sphere of influence of the planet (the Hill sphere) is
smaller than the pressure scale height of the disc. It is easy to see
that for a small enough planet, the upper layers of the disc do not
feel the gravitational pull of the planet, while in the above
two-dimensional theory they fully contribute to the torque. A torque
formula that accounts for this effect reads \cite{tanaka}
\begin{equation}
\Gamma_\mathrm{3D,lin}=-\left(1.364+0.541\alpha\right)
\frac{q^2}{h^2}\Sp\rp^4 \op^2.
\label{eq3D}
\end{equation} 
Note that the difference between the 2D and the 3D result strongly
depends on the surface density profile. In favourable cases, migration
speeds can be reduced by a factor of a few. In the 2D numerical
simulations presented below, a gravitational softening parameter $b$ is
used to account for 3D effects in an approximate way:
\begin{equation}
\Phi_\mathrm{p} =
-\frac{\mathrm{G}M_\mathrm{p}}{\sqrt{|r-\rp|^2+\rp^2b^2}}.
\end{equation}
It is expected that for $b\sim h$, 3D effects on the Lindblad torque
can be captured approximately. The same may not be true for the
non-linear torque discussed below. In \cite{bate}, equation \ref{eq3D}
has been successfully compared to 3D isothermal numerical
simulations. Note, however, that their adopted density profile allows
for weak corotation torques only.
 
The above analysis was simplified by neglecting magnetic fields,
self-gravity and detailed thermodynamics. Including magnetic
resonances in the linear analysis can reduce the torque significantly
if a toroidal magnetic field is present \cite{terquem}. On the other
hand, self-gravity tends to accelerate Type I migration due to a shift
in resonances \cite{pierens,clement2}. More recently, releasing the
assumption of an isothermal disc has been the subject of intensive
study, since simulations presented by \cite{paard06} which
included radiation transport in the disc suggest that type I migration
may be stopped or reversed in a disc which cools inefficiently.
A linear analysis \cite{clement1} suggested that a linear
effect due to the presence of a radial entropy gradient may reduce or
even reverse migration. However, it was subsequently shown
\cite{paardpap08} that this effect is in fact small and can not
account for the results obtained in \cite{paard06}.

\begin{figure}
\includegraphics[width=0.8\columnwidth]{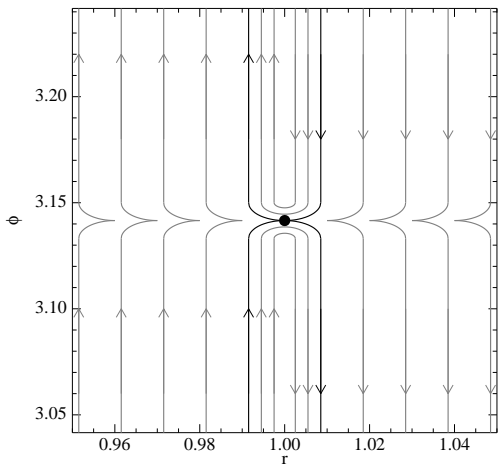}
\caption{Schematic view of the streamline pattern close
  to the planet, which is located at $(r,\varphi)=(1,\pi)$ and is denoted
  by the black circle. Material inside the black streamline (the
  separatrix) executes horseshoe turns on both sides of the planet.} 
\label{fig3}
\end{figure}

\subsection{Beyond the linear model} 

Although intuitively, one would expect that low-mass planets induce
linear perturbations only, recent work has shown that this is not the
case. In a sense, introducing a planet in a disc amounts to a singular
perturbation at corotation, suggesting that the corotation torque can
differ from its linear value. 

The corotation torque is due to material that, on average, corotates
with the planet. Two ways of obtaining the corotation torque can be
found in the literature: one can perform a linear analysis of
perturbed circular orbits around the corotation resonance
\cite{gt79,tanaka} (see above), or, alternatively, one can look at the
expected pattern of the streamlines close to the planet (see figure
\ref{fig3}) and analyse the torque due to material executing 
horseshoe turns \cite{ward91}. We will denote the former result the
\emph{linear corotation torque}, and the latter the \emph{horseshoe
drag}. It is important to stress that there is \emph{no} horseshoe
region in linear theory, since these bends are a genuinely non-linear
phenomenon.    

\subsubsection{Isothermal results}
Considering again two-dimensional, isothermal (or, more general,
barotropic) flow, in which case specific vorticity is conserved along
streamlines, it is easy to see that when executing the turn, material
should change its density. Since it is forced to a different orbit by
the planet, the vorticity of the material changes, and conservation of
vortensity then dictates that there should be an associated change in
density. This change in density is of opposite sign on the different
sides of the planet, hence the planet feels a torque. This torque
depends sensitively on the width of the horseshoe region $\xs$
\cite{ward91}: 
\begin{equation}
\label{eqHorse}
\Gamma_\mathrm{c,hs}=\frac{3}{4}\left(\frac{3}{2}-\alpha\right)\xs^4
\Sp \rp^4 \op^2.
\end{equation}
Note that $\xs$ is in units of $\rp$. Just as in the linear case, the
horseshoe drag is proportional to the radial gradient in specific
vorticity. In the special case of $\alpha=3/2$, the density change
needed to compensate for the change in vorticity exactly amounts to
the background profile, resulting in no torque on the planet.    

Until recently, it has never been clarified how equations \ref{eqCor}
and \ref{eqHorse} are related. There is no reason why the value of
$\xs$, unspecified in the original theory \cite{ward91}, should adjust
itself to match the linear torque. Detailed modeling of the horseshoe
region \cite{horse} shows that $\xs \propto \sqrt{q/h}$, so that the
horseshoe drag scales in exactly the same way as the linear corotation
torque. The only way they differ is in magnitude, and in general the
horseshoe drag is much stronger \cite{drag}.

It is intuitively clear that these two distinct torques can not exist
at the same time: material either follows a perturbed circular orbit
or executes a horseshoe turn. At sufficiently early times after
inserting the planet in the disc, therefore, before any turns can have  
occurred, one would expect linear theory to be valid. Once material
starts to execute horseshoe turns, the linear corotation torque will
get replaced by the non-linear horseshoe drag, which, as remarked
above, will in general be much stronger.    

\begin{figure}
\includegraphics[width=0.8\columnwidth]{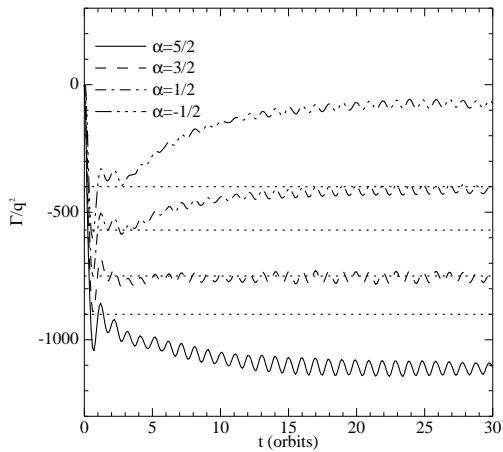}
\caption{Torques on a $q=1.26\cdot 10^{-5}$ planet (4 $\me$ around a
  Solar mass star) embedded in an inviscid disc with $h=0.05$. All
  torques have been normalised to $\Sp=\rp=\op=1$. Results for
  different density profiles are shown, with dotted horizontal lines
  indicating results obtained from linear theory. A gravitational
  softening of $b/h=0.6$ was used.}  
\label{fig4}
\end{figure}

This can be verified using numerical hydrodynamic
simulations. Resolution is crucial, since for low-mass planets $\xs
\ll h \ll 1$, and in order to obtain reliable estimates of the torque,
the horseshoe bends need to be resolved. Simulation results for different values
of $\alpha$ are displayed in figure \ref{fig4}, showing the total
torque on the planet as a function of time. Overplotted are the
results from solving the linear equations (dotted lines). It is clear
that only in the case where $\alpha=3/2$ the non-linear hydrodynamic
calculations match linear theory. In this special case, the corotation
torque vanishes. For all other values of $\alpha$, a departure from
linear theory can be seen, pointing to a stronger corotation torque
than expected from linear theory. However, the total torque is
consistent with the linear corotation torque replaced by the stronger
horseshoe drag, showing that indeed non-linear effects play an
important role in the corotation region. 

Another important aspect is the time scale to set up the torque. Since
the linear corotation torque has no definite scale, it takes
approximately one orbit to develop. The horseshoe drag, on the other
hand, does have an intrinsic scale, namely $\xs$. An associated time
scale is the libration time, which is basically the time it takes for
a fluid element, initially at a distance $\xs$ from the orbit of the
planet, to complete one whole horseshoe orbit:
\begin{equation}
\tau_\mathrm{lib}=\frac{8\pi}{3\xs}\op^{-1}.
\end{equation}
It takes a fraction of the libration time for the horseshoe drag to
develop, which is therefore different for different planet masses
since $\xs \propto \sqrt{q}$. The transition between the linear result
(obtained after approximately 1-2 orbits), and the more slowly
developing non-linear horseshoe drag can be clearly seen in figure
\ref{fig4}. 

\subsubsection{Extension to non-isothermal discs}
These simple models of Type I migration lack a crucial ingredient: a
proper treatment of the disc thermodynamics. Most previous simulations
were indeed done in the locally isothermal limit, where the
temperature of the disc is a prescribed and fixed function of
radius. Such a model would apply if any excess energy could be
radiated away very efficiently. However, in discs that can not cool
efficiently (i.e. are optically thick), the presence of a radial
entropy gradient leads to strong coorbital torques that can reverse
the direction of migration \cite{paard08}. This can again be
understood in terms of the horseshoe drag. When the gas is not able to
cool (the adiabatic limit), entropy is conserved along streamlines. In
a similar way as the vortensity gradient led to the barotropic
horseshoe drag, a radial entropy gradient leads to a strong horseshoe
drag related to density changes as the fluid tries to conserve its
entropy when making the horseshoe turn \cite{paardpap08}. When the
entropy decreases outward, this horseshoe drag gives a positive
contribution to the torque. 

\begin{figure}
\includegraphics[width=0.8\columnwidth]{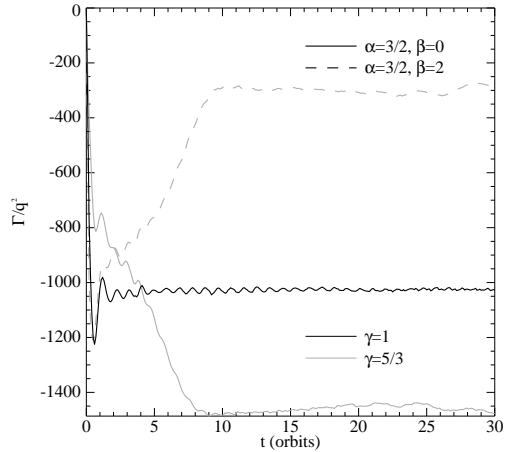}
\caption{Torques on a $q=1.26\cdot 10^{-5}$ planet (4 $\me$ around a
  Solar mass star) embedded in an inviscid disc with $h=0.05$. All
  torques have been normalised to $\Sp=\rp=\op=1$. The black line
  indicates the isothermal result, grey curves results from adiabatic
  calculations for two different temperature profiles. A gravitational
  softening of $b/h=0.4$ was used.}  
\label{fig5}
\end{figure}

This is illustrated in figure \ref{fig5}, where we compare adiabatic
and isothermal simulations with $\alpha=3/2$ for different temperature
profiles (parametrised by $\beta = -d\log T/d\log r$). An adiabatic
exponent $\gamma=5/3$ was chosen. Due to the isentropic nature of
linear waves, the density of the wakes is lowered by a factor $\gamma$
compared to isothermal simulations, which makes the Lindblad torque a
factor $\gamma$ smaller. Note that the density profile is such that
the barotropic corotation torque vanishes. The entropy follows a power
law, initially, with index $-\xi=-\beta+(\gamma-1)\alpha$. Therefore,
the isothermal disc ($\beta=0$) with no vortensity gradient
($\alpha=3/2$) does have an entropy gradient ($\xi=-1$), giving rise
to a strong horseshoe drag. This is clear from the grey solid line,
which shows a large swing between 2 and 10 orbits as a result. When we
modify the temperature structure to reverse the entropy gradient, a
corresponding opposite swing is observed (dashed grey line in figure
\ref{fig5}). Even without an additional contribution from a vortensity
gradient, the torque is reduced by a factor of 3 compared to the
linear value. 

\begin{figure}
\includegraphics[width=0.8\columnwidth]{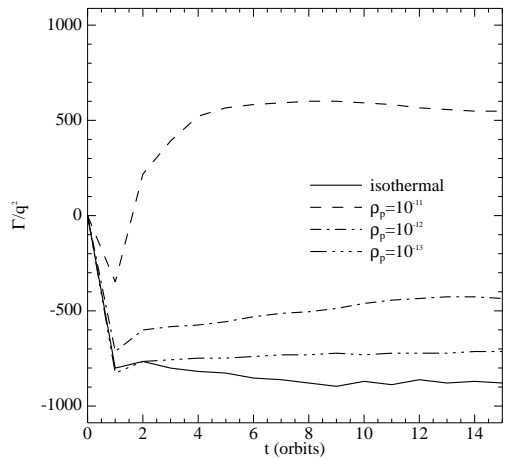}
\caption{Torques on a $q=1.26\cdot 10^{-5}$ planet (4 $\me$ around a
  Solar mass star) embedded in a 3D inviscid disc with $h=0.05$. All
  torques have been normalised to $\Sp=\rp=\op=1$. Results for
  different midplane densities are shown, and the differences between
  the curves are a result of the corresponding change in opacity,
  which affects the cooling properties of the disc. 
  }  
\label{fig6}
\end{figure}

Finally, we release the two-dimensional approximation, and enable the
disc to cool through radiation. The resulting 3D
radiation-hydrodynamical models \cite{paard06} are the most realistic 
simulations of low-mass planet migration so far. In figure \ref{fig6},
we show the results for $\alpha=1/2$ and $\beta=1$. Different midplane
densities $\Rp$ were used, where $\Rp = 10^{-11}$
$\mathrm{g}~\mathrm{cm}^{-3}$ approximately corresponds to the minimum
mass Solar nebula at 5 AU. Note that the torques are still normalised
to $\Sp=1$. Therefore, different densities mainly indicate different
opacities.  

The high density case is almost equivalent to an adiabatic
simulation. This is easy to understand, since the disc is very
optically thick in this case and basically can not cool through
radiation. This again leads to a positive torque, and outward
migration. Lowering the density leads to more efficient cooling, and
one would expect to return to the isothermal result when the opacity
is low enough. This is indeed the case: upon decreasing the density by
a factor of 10 the torque is already negative, and with another factor
of 10 we can almost reproduce the isothermal result. Therefore, the
entropy-related horseshoe drag is important in the dense inner regions
of protoplanetary discs. This mechanism can then act as a safety net
for low-mass planets, since they stop their inward journey as soon as
they reach the dense inner regions of the disc. 

\begin{figure*}
\centering \includegraphics[width=0.3\linewidth]{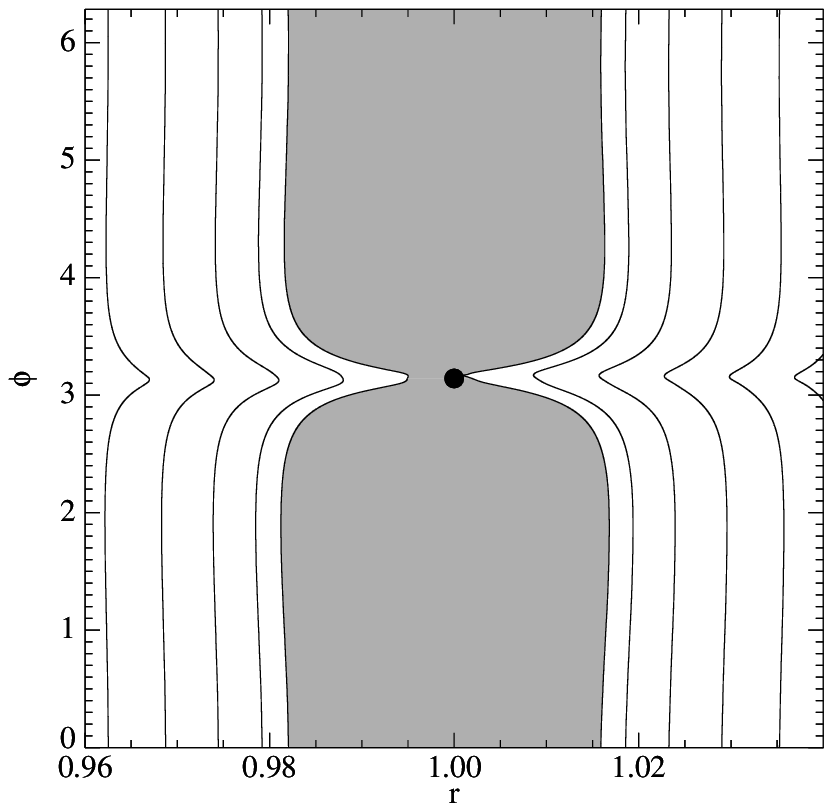}
\centering \includegraphics[width=0.3\linewidth]{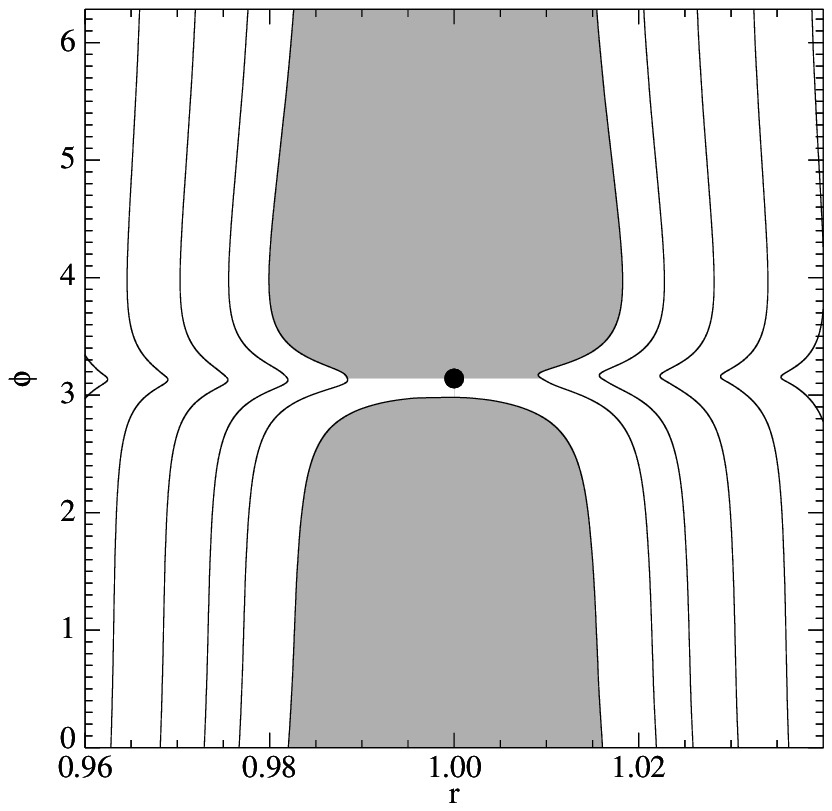}
\centering \includegraphics[width=0.3\linewidth]{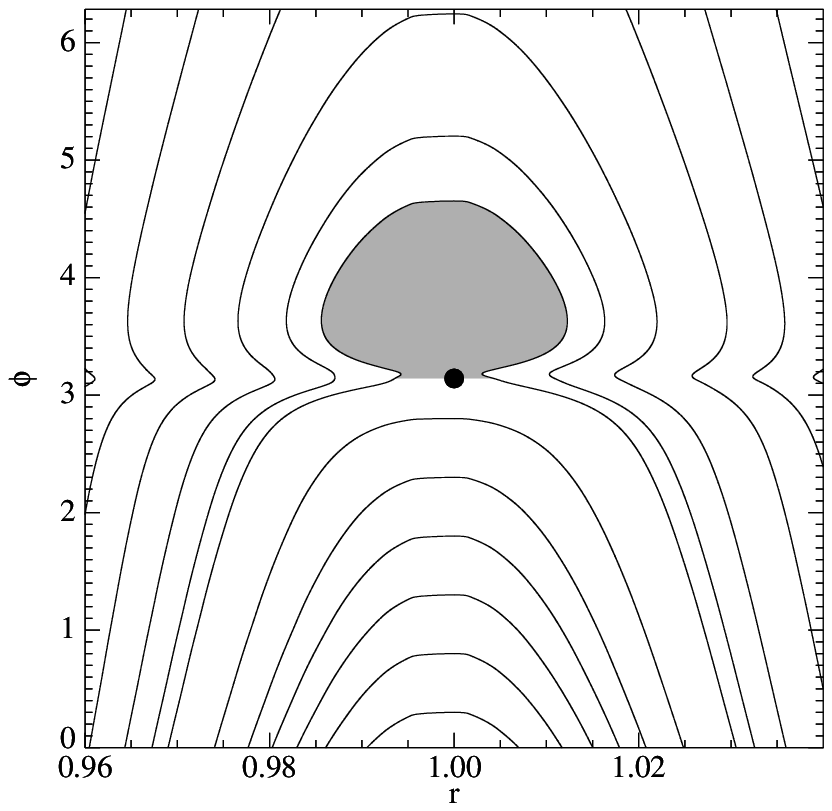}
\caption{Streamlines near the corotation region for different
  migration speeds, in a frame moving with the planet. Left panel: no
  migration, middle panel: slow inward migration, right panel: fast
  inward migration.  
  }  
\label{fig7}
\end{figure*}

\subsubsection{Saturation}
In absence of any diffusive process in the disc, the horseshoe region
is a closed system, and it can therefore only give a finite amount of
angular momentum to the planet. Phase mixing leads to a flattening of
the vortensity/entropy profiles, which results in a vanishing torque
after the gas has made several horseshoe turns. In other words, the
corotation torque saturates \cite{ogilvie}. The horseshoe region needs
to be fed fresh material in order to sustain the torque. This can be
done through viscous diffusion in the isothermal case \cite{masset01},
and it has been observed that by including a small kinematic viscosity
and heat diffusion the torque can be sustained \cite{paardpap08}. This
has also been shown with radiative cooling instead of heat diffusion
in \cite{kleycrida}.

\subsection{Outstanding issues}
The non-linear torque strongly depends on the width of the horseshoe
region $\xs$. It is therefore of critical importance to have a good
physical understanding of what determines this width when
three-dimensional effects are taken into account.

It is not clear yet how the entropy and vortensity related torques
work together. Vortensity is not conserved in general along
streamlines in 3D non-barotropic discs, but nevertheless there appears
to be a contribution from the radial gradient in vortensity to the
horseshoe drag in this case. A proper understanding of these processes
will lead to an improved migration law for low-mass planets.  

\subsection{Concluding remarks}
Type I planetary migration is not as well-understood as previously
thought. Even in the simple isothermal case, strong deviations from
linear theory occur whenever the disc has a radial gradient in
specific vorticity. In other words, the corotation torque is always
non-linear. Therefore, one should be careful in applying the widely
used linear torque formula presented in \cite{tanaka}. Even more so
for non-isothermal discs, where a radial entropy gradient can
contribute to the non-linear torque to the effect of reversing the
direction of migration. In order for the torque to be sustained, some
diffusion of energy and momentum is needed. 

\section{Type III migration}
\label{typeIII}
The most recently discovered migration mechanism, first referred to as
runaway migration \cite{maspap}, relies on strong corotation torques
induced by the radial movement of the planet itself as it
migrates. Later, it has become known as Type III migration
\cite[e.g.][]{adam1}. 

\subsection{Theory}
We will return to the simple case of a laminar, isothermal
disc. Consider the horseshoe region as depicted in the left panel of
figure \ref{fig7}, and for now assume that it has constant specific
vorticity, either through saturation or due to the initial density
profile. The corotation torque therefore vanishes. If we now let the
planet migrate radially, the horseshoe region as viewed in a frame
comoving with the planet changes its shape (middle panel of figure
\ref{fig7}). If the planet is moving inward, as is the case in figure
\ref{fig7}, some fresh material from the inner disc will enter the
horseshoe region, make the turn and move into the outer disc, exerting
a torque on the planet in the process. It is easy to see that this
torque is proportional to the migration speed, and acts to accelerate
the migration. There is therefore a positive feedback loop, and the
possibility of a runaway process.

\begin{figure}
\includegraphics[width=0.8\columnwidth]{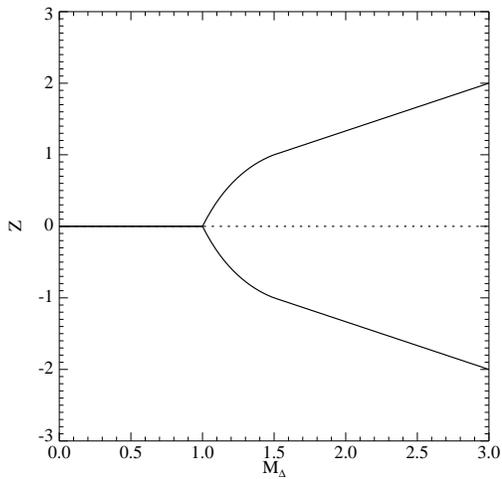}
\caption{Non-dimensional migration rate as a function of the mass
  deficit parameter $M_\Delta$. 
  }  
\label{fig8}
\end{figure}

The mass flow relative to a planet migrating at a rate $\dot\rp$ (we assume a
circular orbit, so that the semimajor axis $a=\rp$) is given by
$2\pi\Sigma_\mathrm{s}\dot\rp$, with $\Sigma_\mathrm{s}$ the surface
density at the upstream separatrix. Since we assume the ordinary,
vortensity-related corotation torque to be fully saturated, the only
horseshoe drag on the planet comes from this fresh material executing
a horseshoe turn, giving rise to a torque \cite{pap07}:
\begin{equation}
\Gamma_{c,3}= 2\pi\Sigma_\mathrm{s}\dot\rp\op\rp^3 \xs.
\end{equation} 
The system of interest that is migrating is the planet itself, mass of gas 
bound within its Roche lobe, and the horseshoe region. The drift rate of
this system is given by
\begin{equation}
(\mpl+M_\mathrm{HS}+M_\mathrm{R})\op\rp\dot\rp =
  4\pi\rp^2 \xs\Sigma_\mathrm{s}\op\rp\dot\rp +2\Gamma_\mathrm{L},
\end{equation}
which can be rewritten as
\begin{equation}
\dot \rp=
\frac{4\pi\rp^2 \xs\Sigma_\mathrm{s}-M_\mathrm{HS}}{\mpl}\dot\rp +
\frac{2\Gamma_\mathrm{L}}{\rp\op},
\label{eqdrift}
\end{equation}
where we have redefined $\mpl$ as the mass of the planet plus the mass
within its Roche lobe. Here, $\Gamma_\mathrm{L}$ is the Lindblad
torque, which may deviate from the linear Lindblad torque due to shock
formation of the appearance of a gap around the orbit of the
planetl. The numerator of the first term on the right 
hand side is called the \emph{coorbital mass deficit}
\cite{maspap}. Basically, this is the difference between the mass the
horseshoe region would have in a disc with $\Sigma=\Sigma_\mathrm{s}$
and its actual mass. It is clear that when this coorbital mass deficit 
becomes comparable to $\mpl$, very high migration rates are
possible. However, in this limit the above analysis becomes invalid. A
more careful treatment \cite{pap07,adamthesis} results in:
\begin{equation}
Z - \frac{2}{3}M_\Delta~ \mathrm{sign}(Z)(1-(1-|Z|_{<1})^{3/2}) =
\frac{2\Gamma_\mathrm{L}}{\mpl \op\rp\dot r_\mathrm{p,f}},  
\label{eqtype3}
\end{equation}
where $M_\Delta$ denotes the coorbital mass deficit divided by the
planet mass, 
\begin{equation}
\dot r_\mathrm{p,f}=\frac{3\xs^2}{8\pi}\rp\op
\end{equation}
is the drift rate required for the planet to migrate a distance $\xs$ in one
libration time, and $Z=\dot\rp/\dot r_\mathrm{p,f}$. A quantity with
subscript $<1$ indicates that the minimum of that quantity and 1 is to
be taken. Equation \ref{eqtype3} gives rise to interesting
behavior. Assuming Lindblad torques to be small, solutions to equation
\ref{eqtype3} are plotted in figure \ref{fig8}. For $M_\Delta <1$,
only one solution, $Z=0$, is possible. There is a bifurcation at
$M_\Delta=1$, beyond which there are three possible solutions: steady
inward or outward Type III migration, plus an unstable solution
$Z=0$. Incorporating $\Gamma_\mathrm{L}$ in this picture removes this
symmetry between inward and outward migration \cite{adamthesis},
favouring migration towards the central star, although outward
migration is still possible \cite{adam2}.

\begin{figure}
\includegraphics[width=0.8\columnwidth]{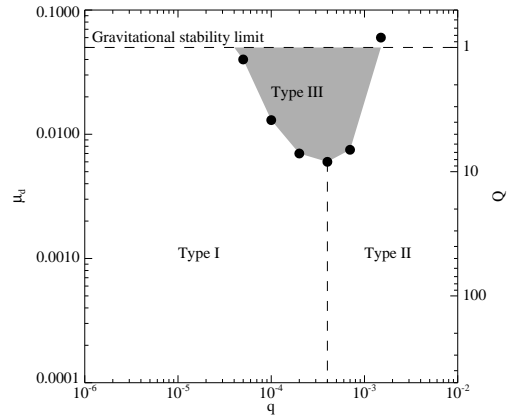}
\caption{The three migration regimes in disc-planet mass parameter
  space, for $h=0.05$. 
  }  
\label{fig9}
\end{figure}

\subsection{Occurence}

\begin{figure*}[ht]
\centering \includegraphics[width=0.8\columnwidth]{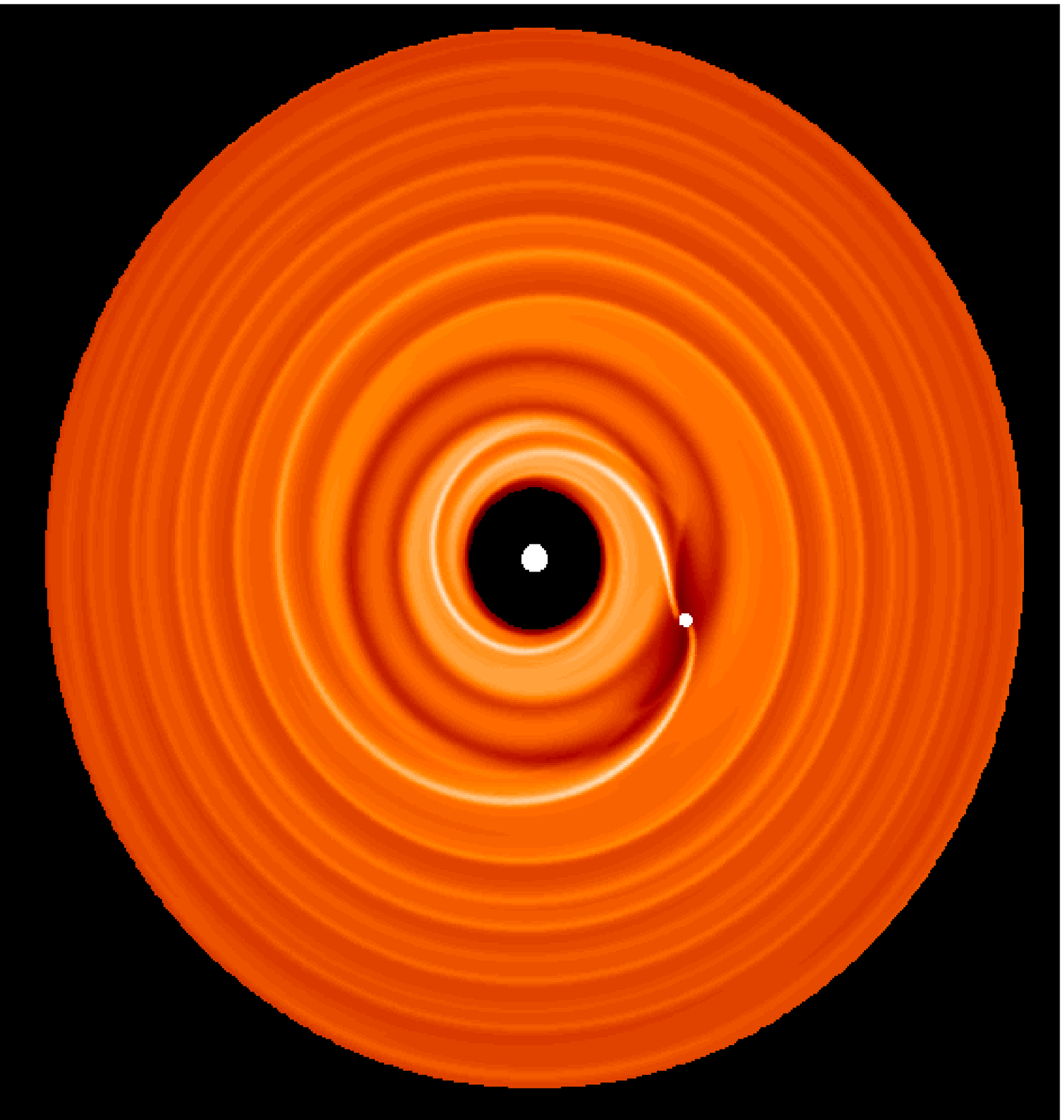}
\centering \includegraphics[width=0.8\columnwidth]{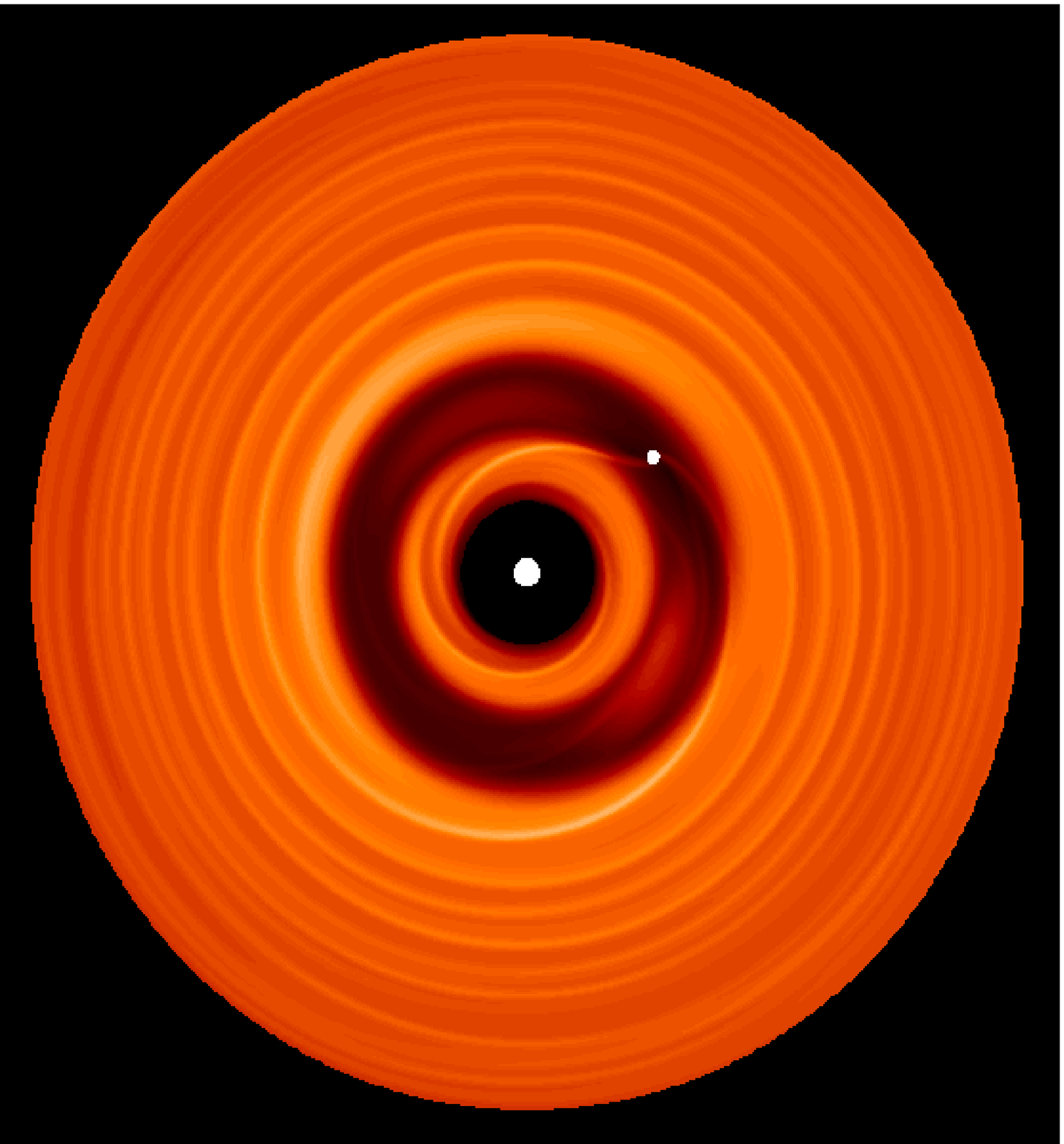}
\caption{\label{Giant-gap}
This figure shows a snapshot of a planet with inital mass of 1 Jupiter mass 
embedded in a protoplanetary disc.
Note the formation of non linear spiral waves and the accompanying gap.}
\end{figure*}

There are different ways in which a coorbital mass deficit can be
obtained. The first possibility is when the planet opens up a (partial)
gap \cite{maspap}. Clearly, in this case there is a difference between
the actual mass of the horseshoe region and the mass of the horseshoe
region with $\Sigma=\Sigma_\mathrm{s}$, hence there is a coorbital
mass deficit. For a high enough disc mass, this deficit can become
comparable to the planet mass. More massive planets require a more
massive disc to enter the Type III migration regime. On the other
hand, low-mass planets do not open up a gap and therefore their
coorbital mass deficit is zero. It turns out that in a typical disc,
Saturn-mass planets are most susceptible to Type III migration
\cite{maspap}. This is depicted in figure \ref{fig9}, where we show
the regime of Type III migration for different disc masses. Here,
$\mu_\mathrm{d}=\pi\rp^2 \Sp /M_*$ is a dimensionless disc mass
parameter, and $Q$ is the corresponding Toomre stability parameter
against fragmentation. For the Minimum Mass Solar Nebula (MMSN) and
$h=0.05$, $Q>10$ inside 10 AU, making Type III migration
unlikely. However, the MMSN is effectively a lower limit on the true
disc mass profile, and for massive discs Type III migration may be an
important effect. 

Another way for a coorbital mass deficit to arise is when the
background disc has a sharp edge. This was studied for Jupiter-mass
planets ($q=0.001$) in \cite{adam1,adam2}. Such a coorbital mass
deficit would also affect low-mass planets, but this has not been
studied so far. 

Type III migration is a numerically challenging problem, since it
strongly depends on material close to the planet. For example, in the
above analysis we redefined the planet mass to include all material
inside the Roche lobe. If the planet obtains a significant envelope,
torques from within the Roche lobe can be large
\cite{dangelo05} and oppose Type III migration. However, since
self-gravity is neglected, this material has effectively no inertia,
making this calculation inconsistent \cite{pap07}.  

To make further progress and deal with such a massive envelope, it is
necessary to incorporate effects of self-gravity, at least in an
approximate way \cite{adam0}. Moreover, to avoid an artificially large
mass build-up inside the Roche lobe due to the isothermal
approximation, a temperature profile for the envelope needs to be
supplied. Only then can simulations of Type III migration, including
torques from inside the Roche lobe, be shown to be converged
\cite{adam0}.  

Type III migration can be very fast. The typical time scale is $\rp/\dot
r_\mathrm{p,f}=2/3\xs^2$ orbits. For $\xs=h$, appropriate for
Saturn-mass planets \cite{horse}, the migration timescale is
approximately 200 orbits, or 2360 year for a planet starting at $5.2$
AU. However, it is not clear at this point how long the coorbital mass
deficit can be retained. For example, unless the surface density
profile decreases steeply with radius, the coorbital mass deficit will
naturally decrease as the planet moves inward because of the changing
factor $\rp^2\Sigma_\mathrm{s}$ in equation \ref{eqdrift}. Outward
Type III migration does not suffer from this effect, but simulations
show that in this case the planet acquires a very massive envelope
that pulls it into the Type II regime \cite{adam2}. Therefore, it
seems that Type III migration is of limited importance for high-mass
planets, changing the semimajor axis by a factor of a few
\cite{adam1,adam2}.    

\section{Type II migration in viscous, laminar discs}
\label{typeII}
\begin{figure*}[ht]
\centering \includegraphics[width=1.0\columnwidth]{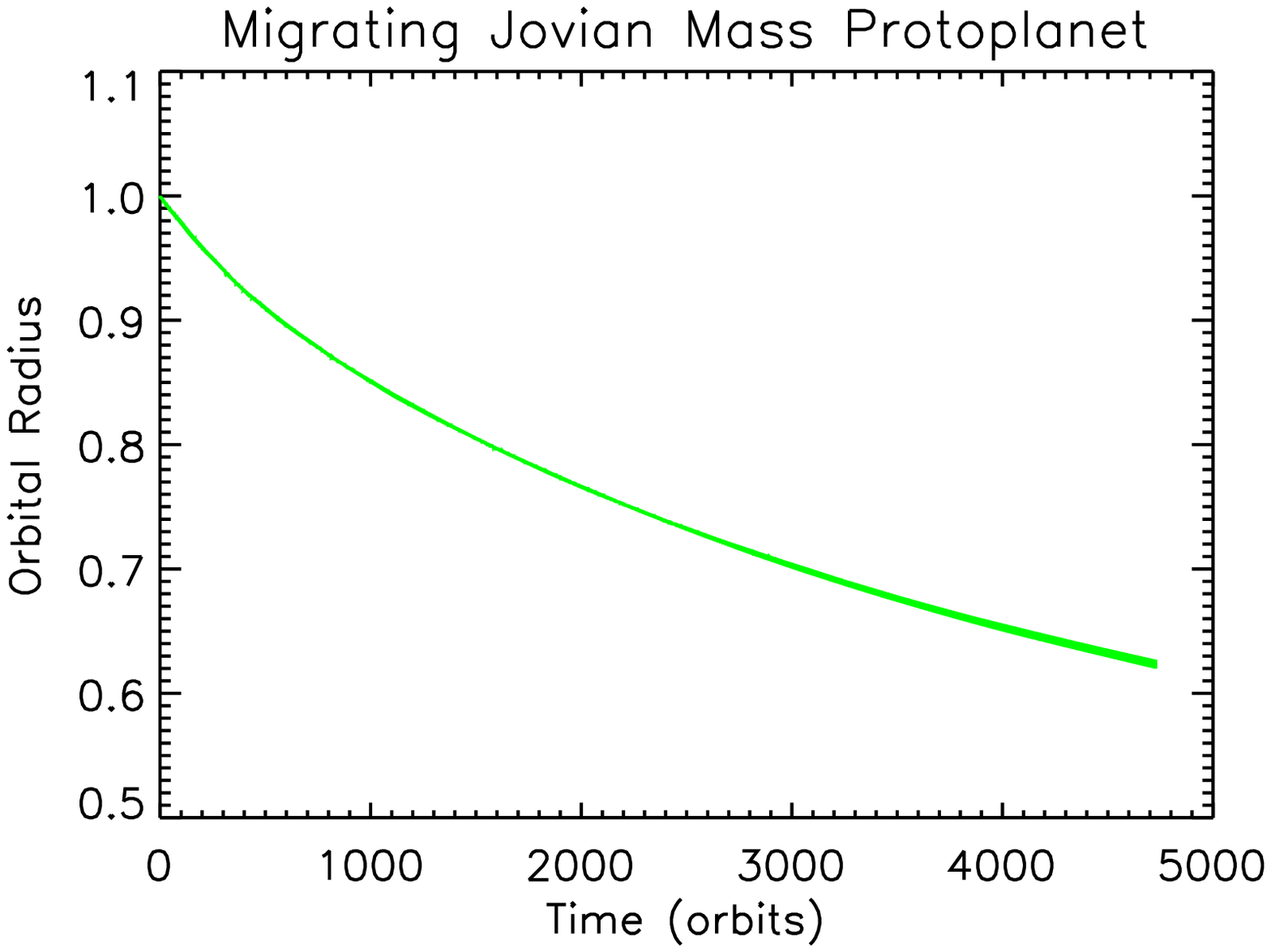}
\centering \includegraphics[width=1.0\columnwidth]{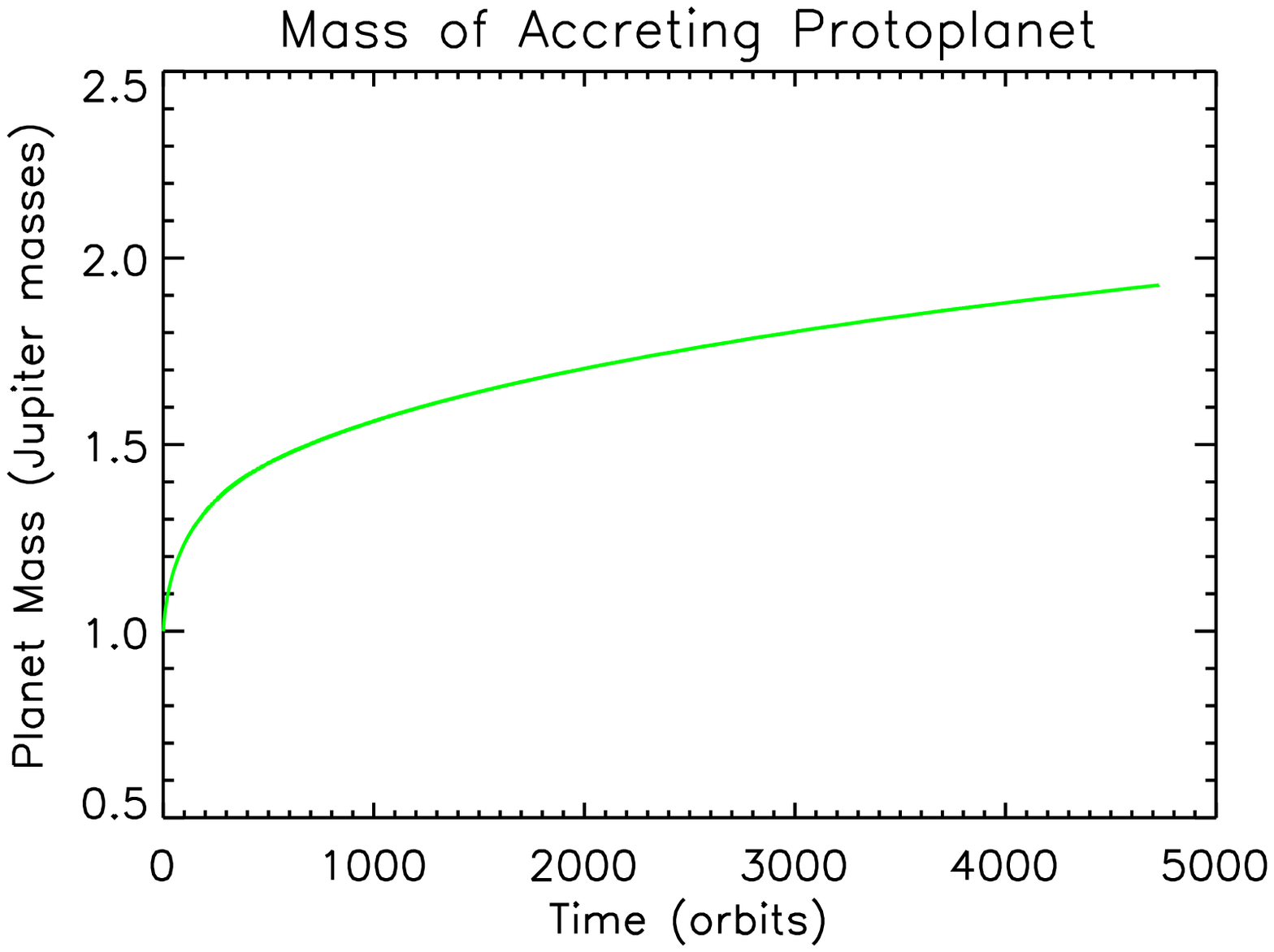}
\caption{\label{Giant-migrate}
The left panel shows the orbital radius of giant planet
as it migrates in the disc model described in the text.
The right panel shows the mass of the planet as it concurrently accretes
gas from the disc.}
\end{figure*}

When a planet grows in mass, the spiral density waves excited by the
planet in the disc can no longer be treated as linear perturbations
in the manner described in section~\ref{typeI}.
The planetary wake becomes a shock-wave in the vicinity of
the Lindblad resonances where it is launched.
Shock dissipation, as well as the action of viscosity,
leads to the deposition of angular momentum associated with the
spiral wave locally in the disc.
Material is thus pushed away from the location of the planet on either
side of its orbital location, and an annular gap begins to form, centred on
the planet semimajor axis. The equilibrium structure and width of the gap
is determined by a balance between the gap-closing viscous and pressure forces
and gap-opening gravitational torques \cite[e.g.][]{cridaetal2006}.

In order for a gap to form and be maintained, two basic conditions must
be satisfied. The first is that the disc response, {\em via} the 
launching of spiral wakes, should be non linear, such that gravitational forces
due to the planet overwhelm pressure forces in the neighbourhood of
the planet. This condition for non linearity is equivalent to the planet
Hill sphere radius exceeding the disc vertical scale height, and may be
expressed mathematically as:
\begin{equation} a\left({\mpl \over 3 M_*}\right)^{1/3} > H. \label{Thc}
\end{equation}
where $a$ is the planet semimajor axis.
The second condition is that angular momentum transport by viscous stresses
in the disc be smaller than planetary tidal forces. This can be expressed
approximately as
\begin{equation} {\mpl \over M_*} > \frac{40 a^2 \Omega}{\nu}. 
\label{viscc}
\end{equation}
where $\nu$ is the kinematic viscosity.
For more discussion about these aspects of gap opening, see the following
papers \cite{linpap1993,bryden99,nelsonetal2000,cridaetal2006}.

For standard disc parameters, $H/r =0.05,$
and $\nu/(a^2\Omega) = 10^{-5},$ equation~(\ref{Thc}) gives
$\mpl/M_* > 3.75\times 10^{-4},$ while 
equation~(\ref{viscc}) gives  $\mpl/M_* >  4\times 10^{-4}.$
The estimates are in reasonable agreement with the results
of numerical simulations of gap opening by \cite{bryden99} and \cite{kley1999}.
Accordingly, for typical protoplanetary disk parameters, we can expect that a
planet with a mass exceeding that of Saturn will begin to open a noticeable gap.

Because the disc response to a massive embedded protoplanet is
non linear, the study of the orbital evolution and gas accretion
by such a planet has been the subject of extensive study {\it via}
numerical simulations. Both 2D \cite[e.g][]{bryden99,kley1999,nelsonetal2000,
dangelo2002}  
and 3D simulations \cite{kley2001,dangelo2003}
have been performed, and overall the results
found in these studies are in good agreement. Differences obviously arise
in the details of the gas flow near the planet when comparing 2D and 3D
simulations, but the global properties of the models are otherwise
similar as the gas is pushed away from the planet vicinity by the process
of gap opening.

Two snapshots from a 2D simulation of a disc
with $H/r=0.05$ and $\nu/(a^2\Omega) = 10^{-5}$ are shown in 
figure~\ref{Giant-gap}, where the initial mass of the planet 
$\mpl=1$ M$_\mathrm{Jup}$ (where M$_\mathrm{Jup}$ is Jupiter's mass). 
The formation of a deep annular gap in the
vicinity of the planet is apparent. The parameters used in this
simulation are such that both the condition for non linearity and
the condition for tides to overwhelm viscous stresses are
satisfied, as described above.

The evolution of the planet semimajor axis is shown in 
the left panel of figure~\ref{Giant-migrate}. 
As the gap forms and the planet migrates inward, 
the migration rate drops significantly below that which is
predicted for type I migration. Once a clean gap is
formed, the planet orbital evolution is controlled by the
rate at which the outer gap edge viscously evolves, such that
the migration time, $\tau_\mathrm{mig}$, can be approximated by 
\begin{equation}
\tau_\mathrm{mig} \simeq \frac{2\rp^2}{3\nu}. 
\label{typeII-mig}
\end{equation}
For typical disc parameters the time to migrate from a distance 
of $\rp =5$ AU to the central star is 
$\tau_\mathrm{mig} \simeq 2 \times 10^{5}$ years.

As the planet mass increases, or the disc mass decreases, the inertia
of the planet can become important and begin to slow the migration. 
This occurs when the local
disc mass contained within the orbital radius of the planet falls
beneath the planet mass ($\pi \rp^2 \Sigma(r_p) < \mpl$).
In this case the migration rate is controlled by the rate at
which mass can build-up at the edge of the gap such that
the disc mass becomes comparable to the planet mass locally \cite{ivanov1999}.
The slowing of migration is apparent in the left
panel of figure~\ref{Giant-migrate}, and a simple linear extrapolation
of the migration history indicates that the time required for this planet 
to migrate into the central star is $\simeq 3 \times 10^4 $ orbits
(or $\sim 3 \times 10^5$ years for a planet initially located at 5 AU.
The right hand panel of figure~\ref{Giant-migrate} shows the
evolution of the planet mass as it accretes gas from the disc.
The mass accretion rate is slowed by the formation of the gap,
but nonetheless gas is still able to diffuse through the gap onto the planet
allowing it to grow.
Linear extrapolation of the planet mass shown in figure ~\ref{Giant-migrate} 
indicates that the planet will have grown to a mass of 
$\mpl \simeq 3.5 M_\mathrm{Jup}$ by the time it reaches the central star,
such that migration and concurrent gas accretion occur during
the gap opening phase of giant planet formation. 

The simulations for gap forming planets presented in this
section indicate that planets grow substantially through
continued gas accretion as they migrate. This raises an
interesting question about the type of correlation which
expected between planetary semimajor axes and masses.
Taken at face value, a model which predicts that planets
grow as they migrate inward will predict an inverse 
correlation between planetary mass and semimajor axes.
The strength of this correlation will be weakened by variations
in protoplanetary disc masses, planet formation times, disc lifetimes
etc, but if these parameters are peaked around certain typical values
in nature, then the correlation should be apparent in the extrasolar
planet data. The absence of such a correlation may in part be
explained by post-formation interaction within a planetary system,
leading to significant gravitational scattering. The
possibility that significant migration may occur prior to
the last stages of giant planet formation may also be important.
Nonetheless, indirect evidence for the past occurence of 
Type II migration in some extrasolar planetary systems
{\em is} provided by multi-planet systems which display
the 2:1 mean motion resonance (or other resonances of low degree).
Such a configuration is most naturally explained by two giant planets
undergoing convergent migration at rates predicted by Type II migration
theory, as illustrated by the numerical simulations presented
in \cite[e.g.][]{snellgrove2001,crida2008}.

\subsection{Outstanding issues}
It has turned out to be difficult to improve on the
non-self-gravitating, isothermal results in which gas inside the Roche
lobe is considered to be part of the planet \cite{maspap}. Approximate
corrections for self-gravity and non-isothermality are necessary to
make progress on high-mass planets \cite{adam0}. The first fully
self-gravitating calculations indicate that this can have important
effects \cite{zhang}. More work is necessary in this area, also to
completely release the isothermal assumption to obtain a realistic
envelope structure for the planet.

\section{Turbulent protoplanetary discs}
\label{sec:turb}
The majority of calculations examining the interaction between protoplanets
and protoplanetary discs have assumed that the disc is laminar.
The mass accretion rates inferred from observations of young stars,
however, require an anomalous source of viscosity to operate
in these discs.
The most likely source of angular momentum transport
is magnetohydrodynamic turbulence generated
by the magnetorotational instability (MRI) \cite{balbushawley1991}
Numerical simulations have been performed using the local shearing box
approximation \cite[see reference][for a review]{balbushawley1998}
and using global geometry \cite[e.g.][]{papnelson2003}. 
In each of these set-ups, runs have been
conducted using non stratified models in which the vertical
component of gravity is ignored, and in more recent work the
vertical component of gravity has been included in shearing box
\cite{millerstone2000}
and global models \cite[][and references therein]{fromangnelson2006}.
These model indicate
that the nonlinear outcome of the MRI is vigorous turbulence, and dynamo
action. Recent work has indicated that the picture is more
complicated than previously thought, with there being a strong
dependence of the saturated state of the turbulence on the value
of the magnetic Prandtl number \cite{fromangetal2007, lesur2007}.

\begin{figure}[ht]
\centering \includegraphics[width=0.8\linewidth]{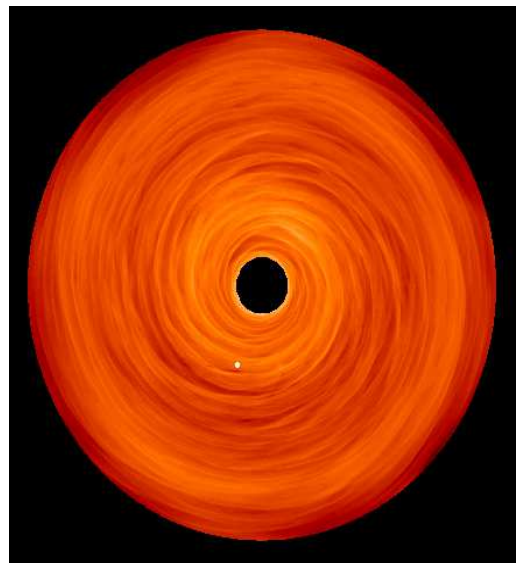}
\caption{\label{turb-fig1} This figure shows a snapshot of a 10 M$_{\oplus}$
protoplanets embedded in the turbulent protoplanetary disc model described in the
text.}
\end{figure}

Prior to the realisation that the strength of the MHD turbulence depends on
the Prandtl number, it was recongnised that non ideal MHD effects
might be important in protoplanetary discs.
The ionisation fraction in cool, dense protoplanetary discs
is expected to be small in the planet forming region
between 0.5 -- 10 AU. Only the surface layers of
the disc, which are exposed to external sources of ionisation such
as X--rays from the central star or cosmic rays, are likely to be sufficiently
ionised to sustain MHD turbulence
\cite[e.g.][]{gammie1996, fromang2002, ilgnernelson2006, ilgnernelson2008}. 
Predictions about the structure and depth of the interior
non turbulent ``dead zone" depend on 
complex chemical reaction networks and the degree of dust grain
depletion, issues which themselves may be influenced by the
process of planet formation {\it via} dust grain growth.

Although there are a number of uncertaintities remaining
about the nature and strength of disc turbulence, it is clear that
an accurate understanding of planet formation and disc-planet
interaction requires the inclusion of turbulence in the models.
Work conducted over the last four or five years 
has examined the interaction between planets of various masses
and turbulent protoplanetary discs, where these studies have usually
simulated explicitly MHD turbulence arising from the MRI.
We now review the results of these studies, and present some more
recent models.

\begin{figure*}[ht]
\centering \includegraphics[width=0.45\linewidth]{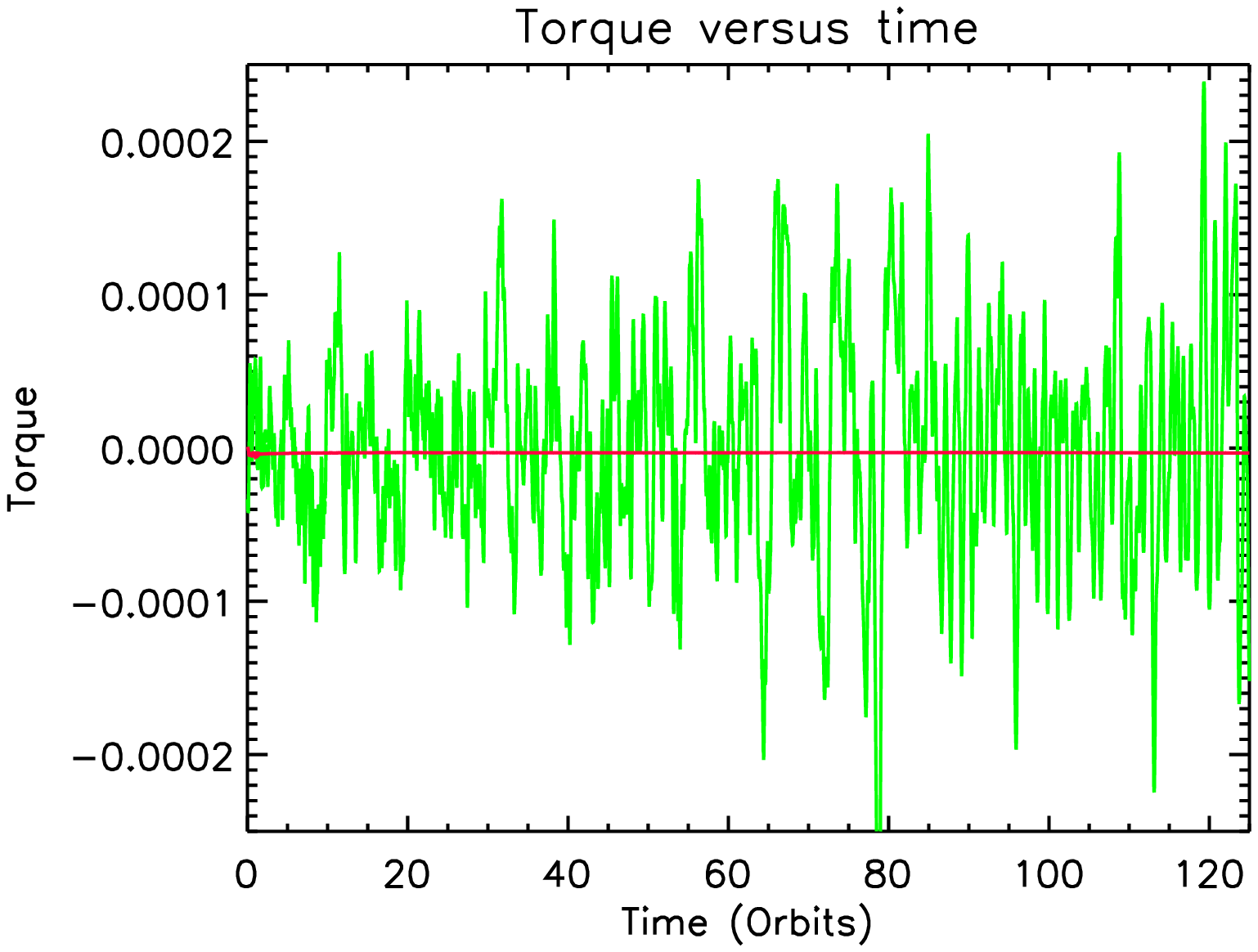}
\centering \includegraphics[width=0.45\linewidth]{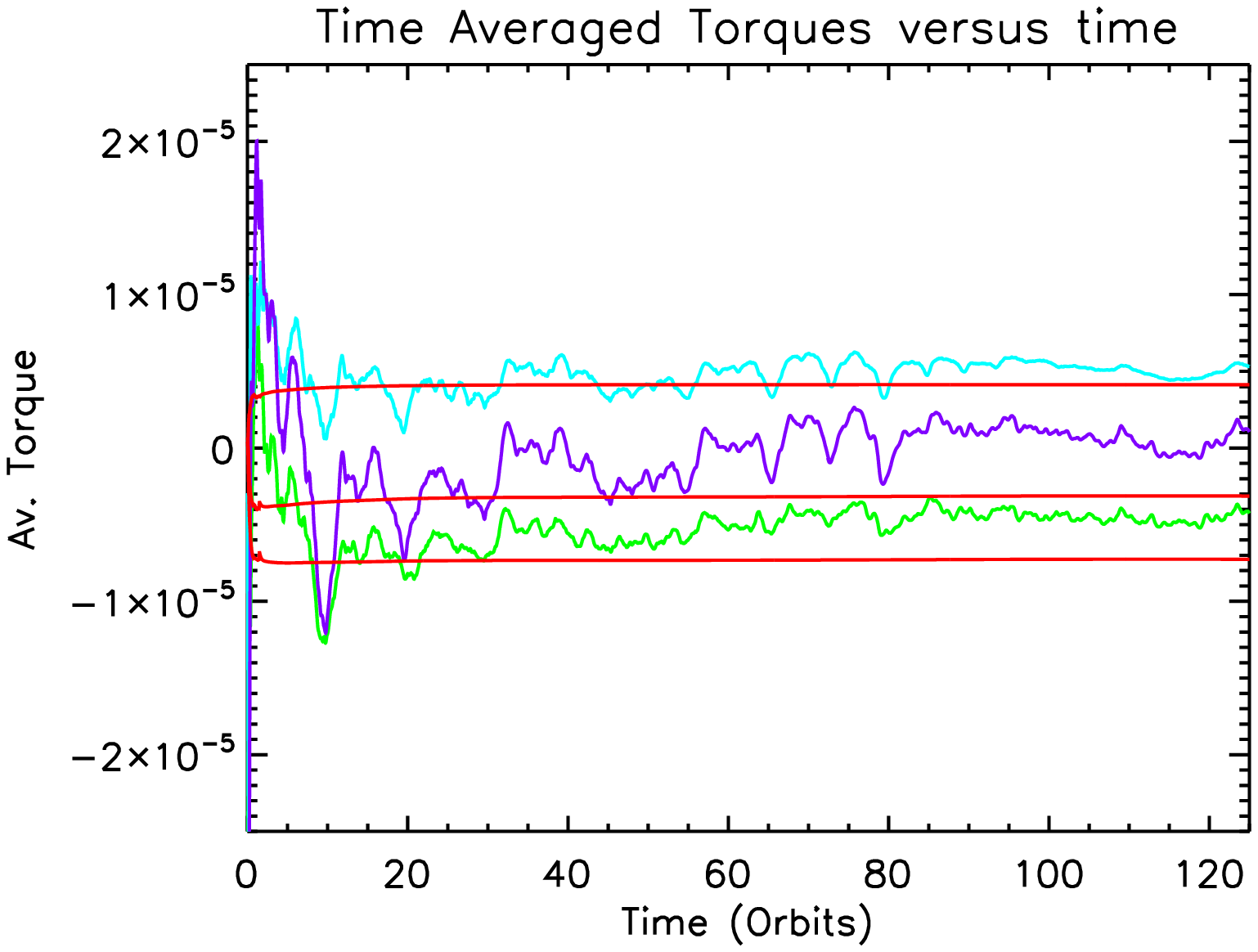}
\caption{\label{turb-fig2} The green line in left panel shows the 
torque experienced
by the planet embedded in the turbulent disc. The red line shows the torque
experienced by a planet embedded in an equivalent laminar disc.
The right panel shows the time averaged values of the torques.
From top to bottom, red lines are the torques from the laminar disc 
calculation corresponding to the disc interior to the planet, 
the total torque, and the disc exterior to the planet.
Similarly the turquoise, purpole and green lines
are from the turbulent disc simulation.
}
\end{figure*}

\begin{figure}[ht]
\centering \includegraphics[width=1.0\linewidth]{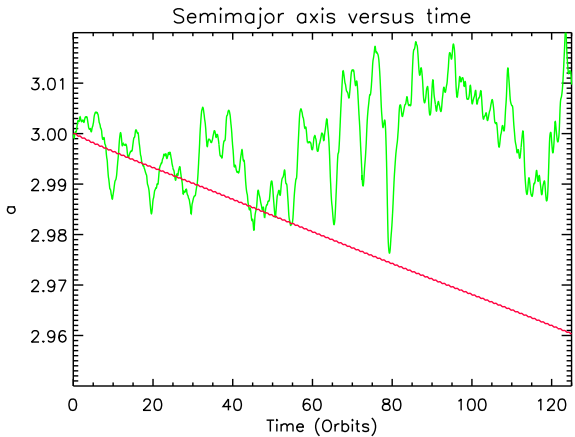}
\caption{\label{turb-fig3} This shows the variation of semimajor
axis with time (measured in planet orbits at $r=3$). The green line shows
the planet evolution in a turbulent disc, the red line shows the
evolution in an equivalent laminar disc.}
\end{figure}

\subsection{Low mass protoplanets in turbulent discs}
\label{low-mass-turb}
The interaction between low mass, non gap forming protoplanets and
turbulent discs has been examined by \cite{nelson04, papetal2004, nelson2005,
laughlin2004}
These calculations show that interaction
between embedded planets and density fluctuations generated by
MHD turbulence can significantly modify type I migration, at least over
time scales equal to the duration of simulations that are currently
feasible ($t\sim 150$ planet orbits). The influence of the turbulence
is to induce a process of `stochastic migration'
rather than the monotonic inward drift expected for planets
in laminar discs. 
The simulations presented in the papers cited above considered 
the interaction of
planets with non stratified (no vertical component of gravity) disc
models (for reasons of computational expense), but the work of
\cite{fromangnelson2006} indicates that the amplitude of midplane density
fluctuations in stratified models is weaker than in their non stratified
counterparts because of magnetic buoyancy effects removing net magnetic
flux from the simulation domain. It is therefore important to consider
the evolution of planets in stratified models.

One such model is presented in figure~\ref{turb-fig1}
which shows a snapshot of the midplane density in a turbulent
disc with an embedded 10 Earth mass planet.
The disc model in this case has aspect ratio $H/r=0.1$, and
was initiated with a ratio of magnetic to thermal gas pressure
$P_\mathrm{g}/P_\mathrm{m} =25$
thoughout the computational domain. As such the model is very similar
to those presented in \cite{fromangnelson2006}. The disc extends from an inner
radius of $1.6$ AU to $16.6$ AU, and has a vertical domain covering a total
of 9 scale heights. The resolution corresponds to 
($N_r$, $N_{\phi}$, $N_{\theta}$) = (464, 280, 1200).
Figure~\ref{turb-fig1} shows the turbulent density fluctuations are of 
higher amplitude than the spiral wakes generated by the planet, as the
planet induced wakes are not visible against the perturbations induced by the
MHD turbulence \cite[see also][]{nelson04, nelson2005}.
Indeed, density fluctuations generated by turbulence in simulations
are typically $\delta \Sigma/\Sigma \simeq 0.15$ -- 0.3, with peak fluctuations
being ${\cal O}(1)$. Thus, on a length scale equivalent to
the disc thickness $H$, density
fluctuations can contain more than an Earth mass in a disc model
a few times more massive than a minimum mass nebula.

\begin{figure*}[ht]
\centering \includegraphics[width=0.45\linewidth]{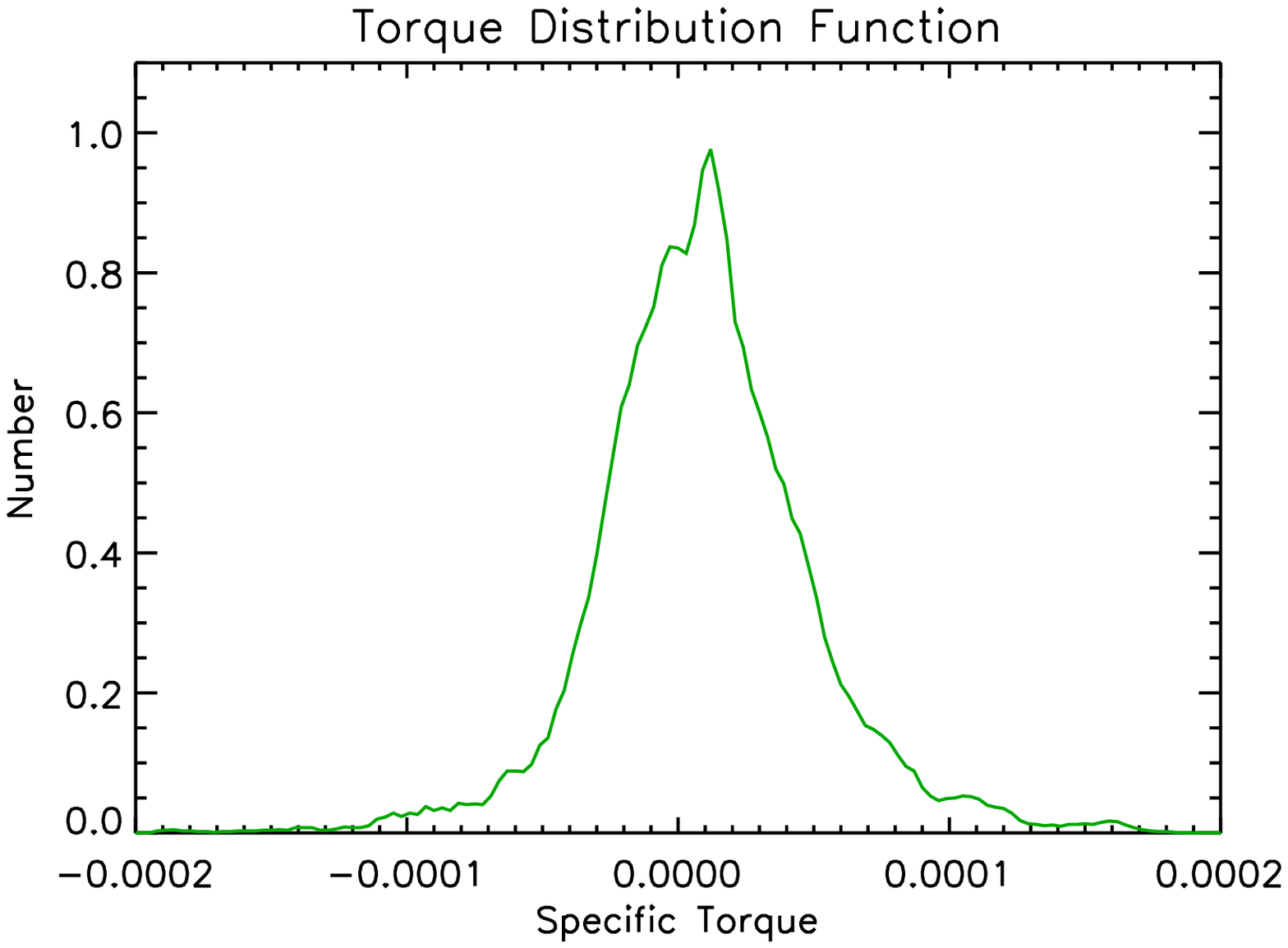}
\centering \includegraphics[width=0.45\linewidth]{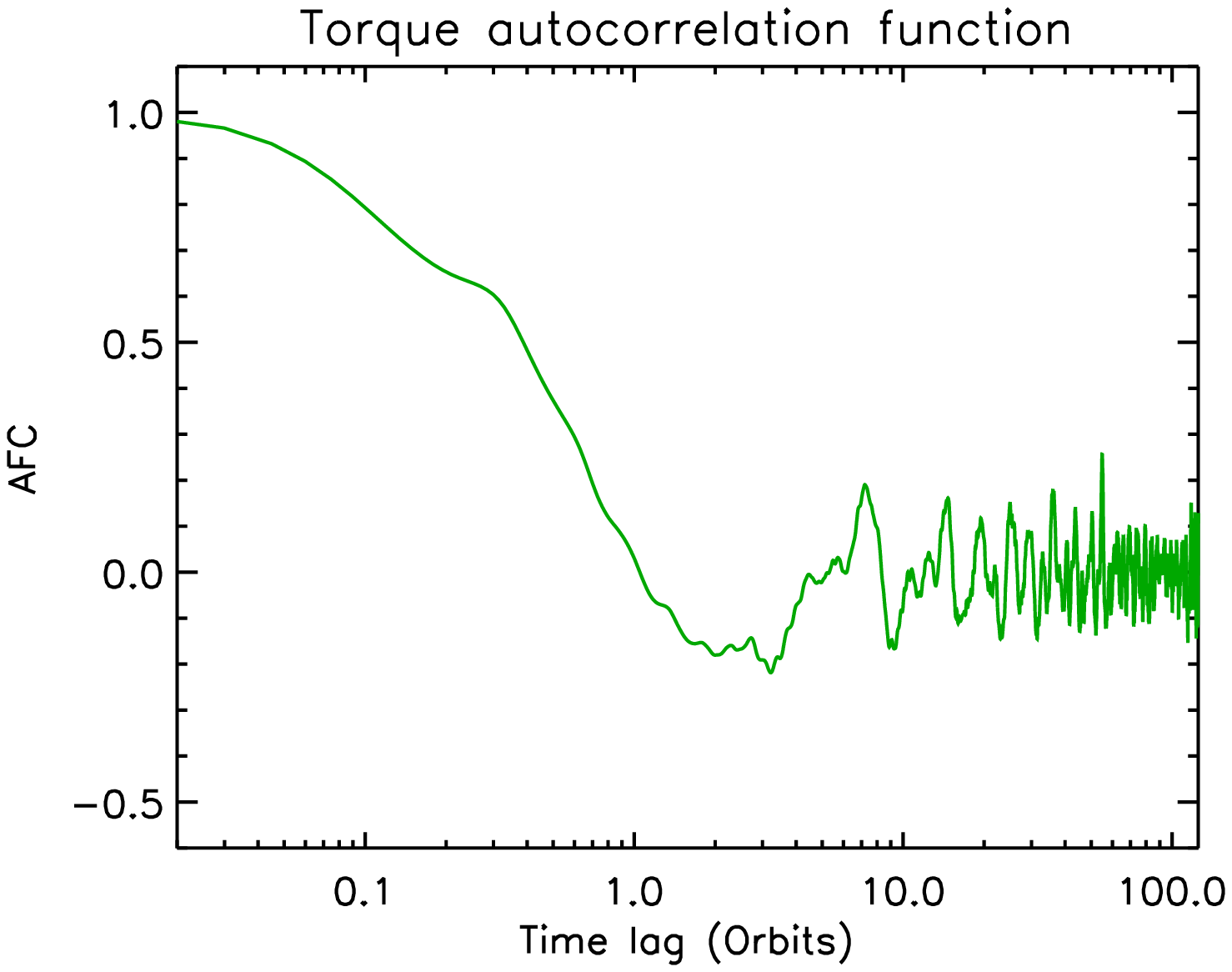}
\caption{\label{turb-fig4} The left panel shows the distribution of
torques experienced by the planet embedded in the turbulent disc.
The right panel shows the autocorrelation function of the torque
plotted in the left panel of Fig.~\ref{turb-fig3}.}
\end{figure*}

The left panel of figure~\ref{turb-fig2} shows the variation in the torque
per unit mass experienced by the planet shown in figure~\ref{turb-fig1}.
The green line shows the torque experienced by a planet in the turbulent
disc model, and the red line shows the torque experienced by the same
planet in an equivalent laminar disc model 
(the torque value here is $\simeq 3 \times 10^{-6}$).
It is clear that the planet
in the turbulent disc experiences strongly varying stochastic torques whose
magnitudes are significantly larger than the expected type I torques.
The right hand panel shows the running time average of the torque
experienced by the planet. The light blue, green and dark blue lines 
correspond to the torque from the outer disc, the inner disc, and the
total torque, respectively. The red lines show the equivalent quatities from
the laminar disc model. It is clear that while the various torque contributions
are of similar magnitudes when comparing the laminar and turbulent models,
the net average torque in the turbulent model is positive, indicating that
the planet has migrated outward slightly over the duration of this simulation.

This expectation is confirmed by figure~\ref{turb-fig3}, which shows
planet semimajor axes for both turbulent (green line) and laminar (red line)
models. It is clear that for the duration of this run the planet in
the turbulent disc has undergone stochastic migration, resulting in it
having moved outward slightly. A key question is whether
the stochastic torques can continue to overcome type I migration over
time scales up to disc life times. A definitive answer
will require very long global simulations, but simple estimates can
be made based on the simulated torques and their statistical properties.

If we assume that the planet experiences type I torques from the
turbulent disc, on top of which are linearly superposed
Gaussian distributed fluctuations
with a characteristic correlation time $\tau_\mathrm{c}$, then we can write
an expression for the time averaged torque experienced by the protoplanet,
${\overline T}$, in the form
\begin{equation}
{\overline T} = <T> + \frac{\sigma_T}{\sqrt{t_\mathrm{tot}}}
\label{time-average}
\end{equation}
where $\sigma_T$ is the standard deviation of the torque amplitude,
$<T>$ is the underlying type I torque,
and $t_\mathrm{tot}$ is the total time elapsed, measured in units of
the torque correlation time $\tau_\mathrm{c}$. In other words we treat 
the calculation as a simple signal-to-noise problem.
Convergence toward the underlying type I value
is expected to begin once the two terms on the right hand side become
equal, as the fluctuating component should begin to average out.
The torque distribution function for this simulation is shown
in the left panel of figure~\ref{turb-fig4}, and the torque
autocorrelation function is shown in the right hand panel.
Expressed in code units, the rms torque value is $\sim 4 \times 10^{-5}$ and
the correlation time is $\tau_mathrm{c} \simeq 1$ planet orbital period.
The type I torque is $<T> \simeq 3.5 \times 10^{-6}$, so the 
time over which type I torques should be apparent is $130$ orbits,
slightly more than the simulation run time.
This prediction suggests that we should expect to see the planet
begin to migrate inward in figure~\ref{turb-fig3}, but this clearly 
is not the case. 

The lack of inward migration in figure~\ref{turb-fig3} may just be a 
consequence of not having run the simulation for long enough,
such that the fluctuating torques have not had time to begin to average
out. The similarity between the run time of the simulation and the
estimate of the time when type I effects should become apparent
suggests that this may be the case. If we look at the right panel of
figure~\ref{turb-fig2}, however, we can see that the averaged
torque due to the disc that lies interior to the planet has converged
to a value which is very close to that obtained in a laminar disc model.
The outer disc torque, however, remains somewhat lower in amplitude,
such that the net torque experienced by the planet is positive over the
simulation run time. It is well known that the stochastic nature of
the disc turbulence, and the associated stresses it generates, can lead to
structuring of the disc radial density profile due to local variations
in the effective viscous stress. This may provide a means of modifying
the balance between inner and outer disc torques such that convergence
to the expected type I value does not happen in practice for an individual
planet in a disc, although it should happen on average for an ensemble
of planets. One potential way in which the underlying type I torques
may not be recovered is if local variations in the viscous stress
lead to undulations in the radial density profile. In this situation
the corotation torque may become important in regions where the disc surface
density increases locally with radius. The effect of this is illustrated
by equation~(\ref{eqCor}), which shows that if the surface density profile
becomes positive then the corotation torque experiences a boost
because $\alpha$ becomes negative. This phenomenon has been dubbed a
`planet trap' by \cite{masset2006}. Over longer viscous time scales
such undulations in the density distribution will be smoothed out,
but it is possible that planets may be able to migrate from one such
density feature to another such that the next migration rate is
substantially reduced. 

Significantly longer simulations are required in order to examine
the long term evolution of low mass planets embedded in 
stratified, global, turbulent disc models and to determine
the efficacy of disc turbulence in modifying type I migration.
Furthermore, in order to build up a statistically significant
picture of the distribution of outcomes, a number of
such simulations will be required. In addition, the issue of
how dead zones affect the stochastic torques also needs to be
examined. It is worth noting that the simulation presented in
this section required approximately 55 days of wall-clock
run time on 256 processing cores. Clearly a full exploration
of this problem remains a very challenging computational task.

\subsection{High mass protoplanets in turbulent discs}
\label{high-mass-turb}
\begin{figure}[ht]
\centering \includegraphics[width=0.8\linewidth]{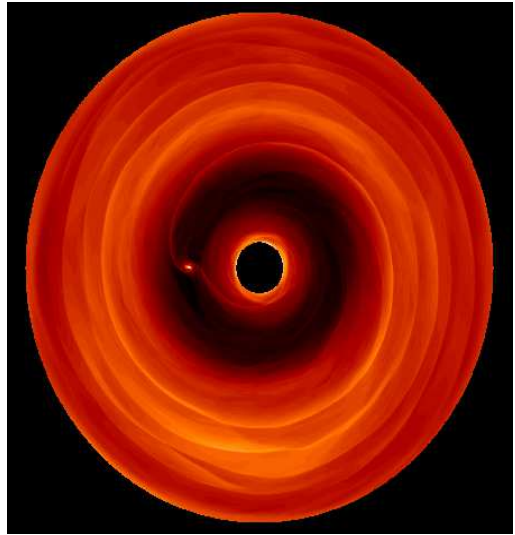}
\caption{\label{turb-fig5} This figure shows a snapshot of the disc
midplane density for a protoplanet with initial mass of $1/3$ M$_{\rm Jup}$ 
embedded in a turbulent disc. This snapshot corresponds to a time when the
planet mass has reached $\simeq 4$ M$_{\rm Jup}$ by accreting gas from
the disc.}
\end{figure}

The interaction of high mass, gap forming planets with
turbulent protoplanetary discs has been considered in a number of papers
\cite{nelsonpap2003, winters2003, papetal2004, nelson04}.
In all these works, the disc model considered was non stratified
such that the vertical component of gravity was neglected.
It is well-known, however, that the flow of material in the vicinity of
a giant planet develops a well-defined three dimensional structure, such that
full 3D simulations are required to examine the details of the disc-planet
interaction.

Figure~\ref{turb-fig5} shows a snapshot of the midplane density for
a turbulent disc with an embedded 4 M$_{\rm Jup}$ protoplanet.
The disc model in this case is the same as that described in 
section~\ref{low-mass-turb} with $H/R=0.1$.
The planet had an initial mass $\mpl = 1/3$ M$_{\rm Jup}$
and was able to accrete gas from the disc. The gas accretion algorithm consists
of the standard procedure of removing a fraction of the gas contained within 
one quarter of the planet Hill sphere at each time step. When magnetic
fields are present in the simulation this procedure can lead to the
formation of a magnetically dominant envelope forming around the planet.
This problem was circumvented by allowing the gas within the planet
Hill sphere to have a finite resistivity such that the field
diffuses from this region as it builds up.

An initial planet mass of $\mpl = 1/3$ M$_{\rm Jup}$ is below the
gap opening mass according to the criterion expresed by equation~(\ref{Thc}).
As the planet accretes mass, however, gas is removed from the
vicinity of the planet, and the growth of the planet mass allows
tides to begin to open a gap once the Hill sphere radius begins to
exceed the local disc scale height.
\begin{figure*}[ht]
\centering \includegraphics[width=0.45\linewidth]{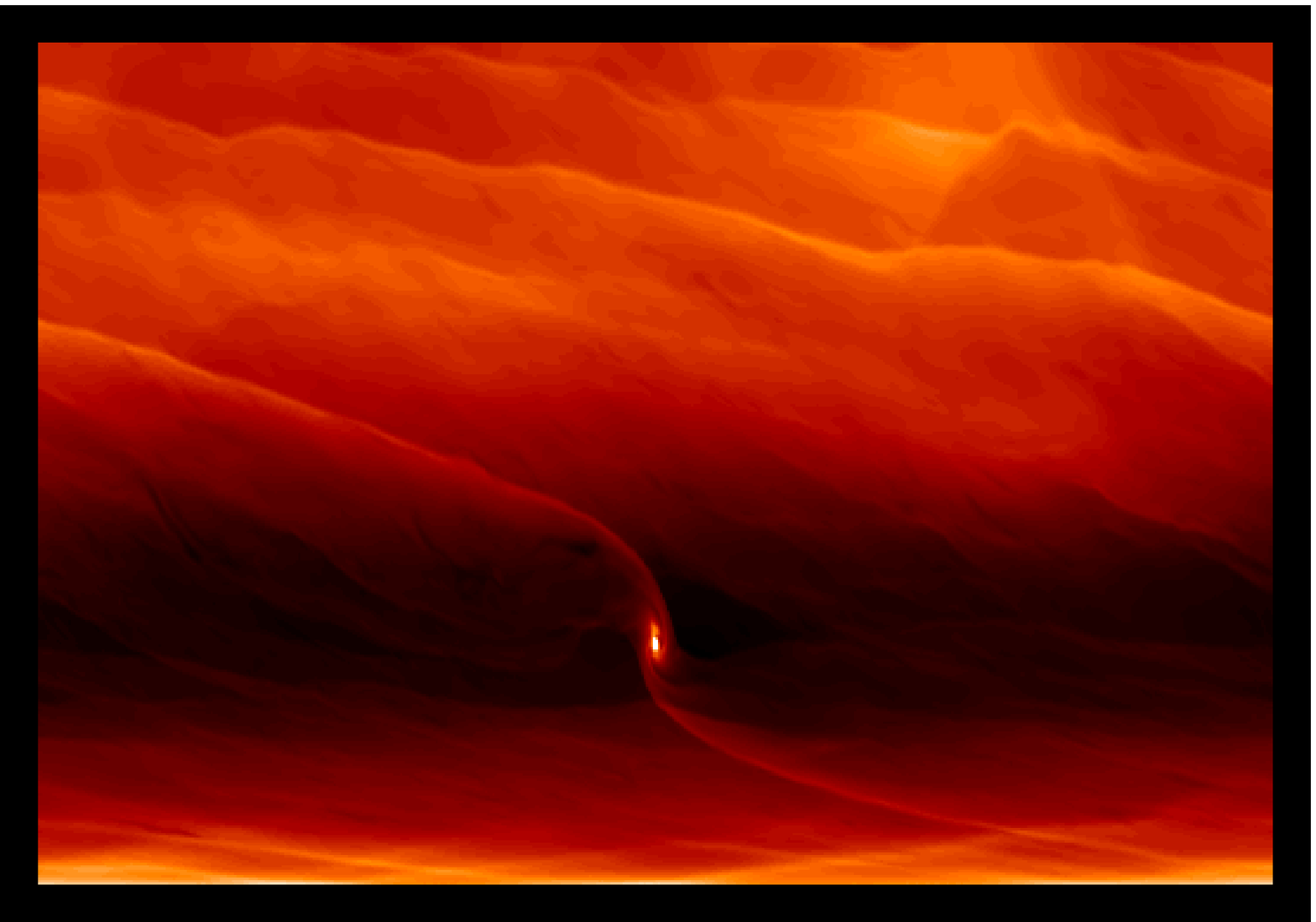}
\centering \includegraphics[width=0.45\linewidth]{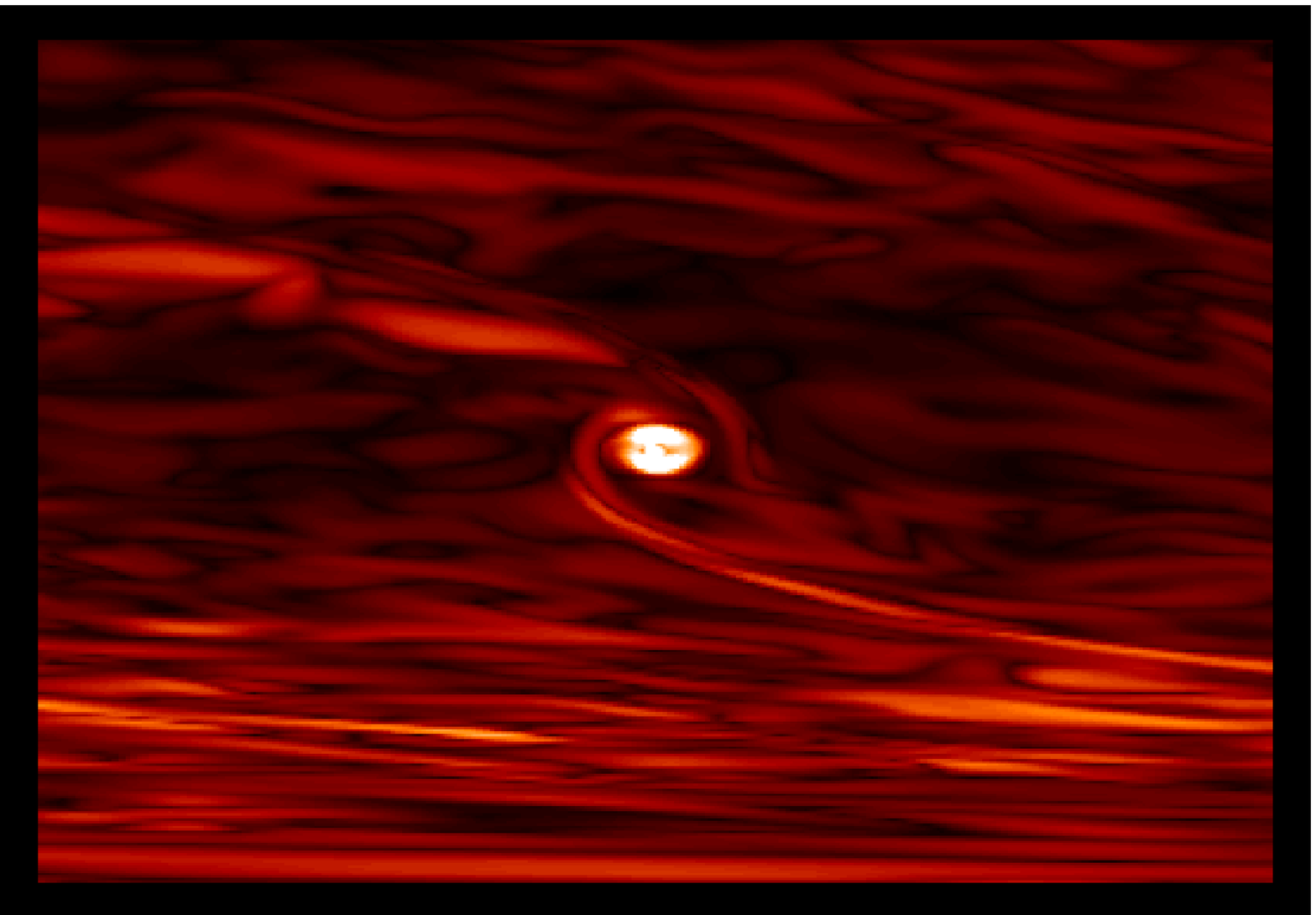}
\caption{\label{turb-fig6} The left panel shows the gas density at
the disc midplane, and the right panel shows contours of the quantity
$B^2$ at the disc midplane.}
\end{figure*}
This is in agreement with \cite{papetal2004},
who considered the transition
from fully embedded to gap forming planets using local shearing box
and global simulations of turbulent discs. These simulations showed
that gap formation begins when the
disc reponse to the planet gravity starts to become non linear.
The combined Maxwell and Ryenolds stresses in simulations with 
zero--net magnetic flux (as considered here)
typically give rises to an effective viscous stress parameter
$\alpha_\mathrm{visc} \simeq 5 \times 10^{-3}$, such that the viscous
criterion for gap formation is satisfied when the criterion for non
linear disc response is satisfied. Here $\alpha_\mathrm{visc}$
is defined to be 
$\alpha_\mathrm{visc} = \langle T \rangle / \langle P \rangle$,
where $T$ is the total (Maxwell plus Reynolds) stress, $P$ is
the thermal pressure, and angled brackets represent vertical
and azimuthal averages.

Global simulations allow the net torque on the planet due to the disc
to be calculated and hence the migration time to be estimated.
Simulations presented in \cite{nelsonpap2003, nelson04} for
massive planets indicate migration times of $\sim 10^5$ yr, in line
with expections for type II migration. The left panel of 
figure~\ref{turb-fig6} shows the torques per unit mass
experienced by the planet due to the disc as a function of time.
Although gap formation has not reached a state of completion
(this is found to require approximately 400 planet orbits
in laminar disc runs such as those presented in section~\ref{typeII}), such
that the torques have not reached their final values, this figure
can be used to estimate a lower limit on the migration time. 
We express this as $\tau_\mathrm{mig} = j/(dj/dt)$, where $j$
is the specific angular momentum of the planet.
The specific torque at $t=100$ orbits is $dj/dt=2 \times 10^{-5}$,
(expressed in code units). The planet specific angular momentum $j=\sqrt{3}$
such that the migration time corresponds to $\tau_\mathrm{mig}=3
\times 10^4$ years. 
Completion of gap formation will reduce the torques experienced
by the planet considerably as the gas is pushed away from
its vicinity by tides, causing the final migration time to significantly exceed
this value. But, as noted in section~\ref{low-mass-turb}, 
the run times associated with these full 3D stratified disc models only
allow full 3D magnetised and stratified disc simulations to
run for $\simeq 100$ orbits at the present time.

\begin{figure*}[ht]
\centering \includegraphics[width=0.45\linewidth]{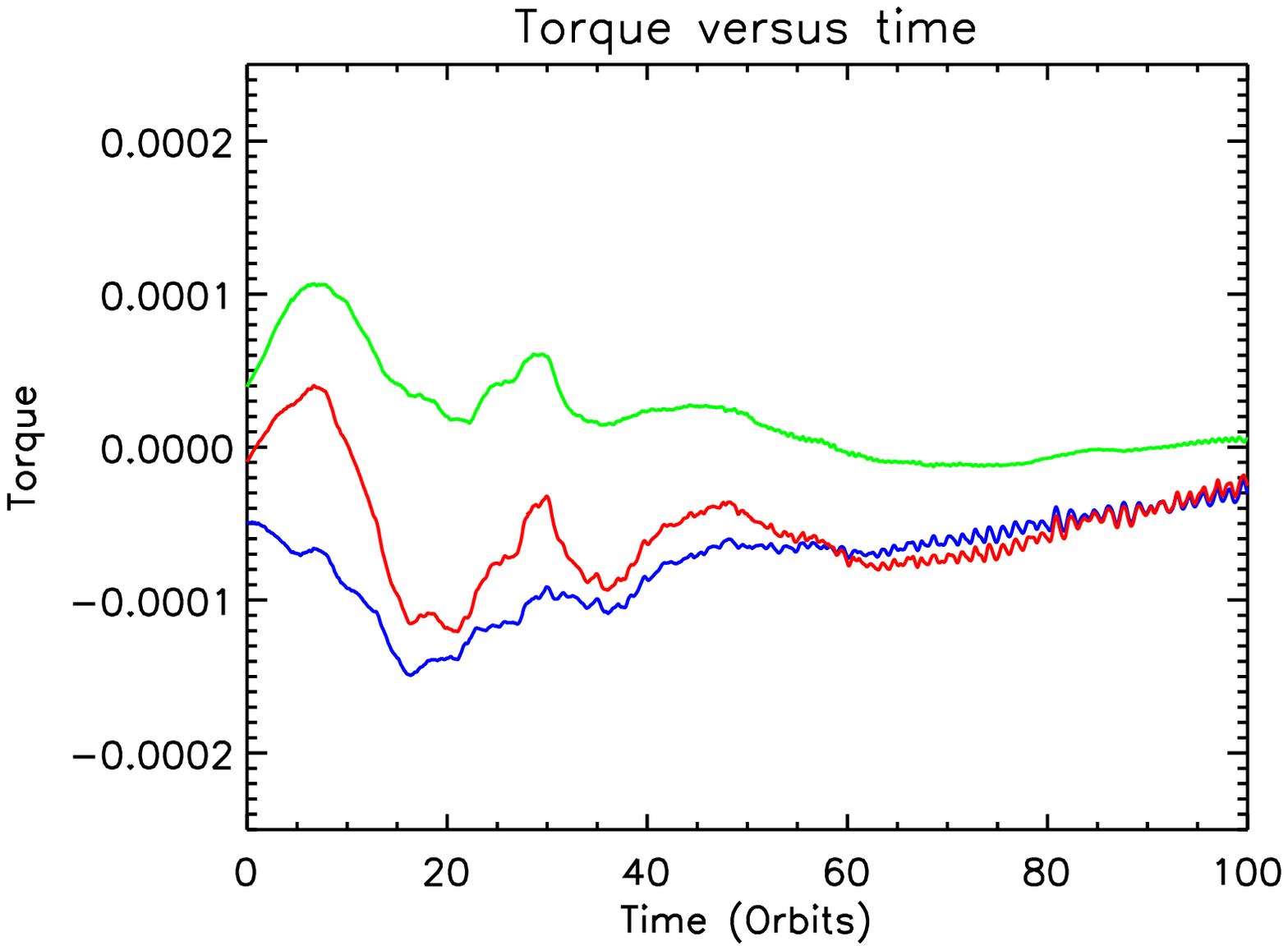}
\centering \includegraphics[width=0.45\linewidth]{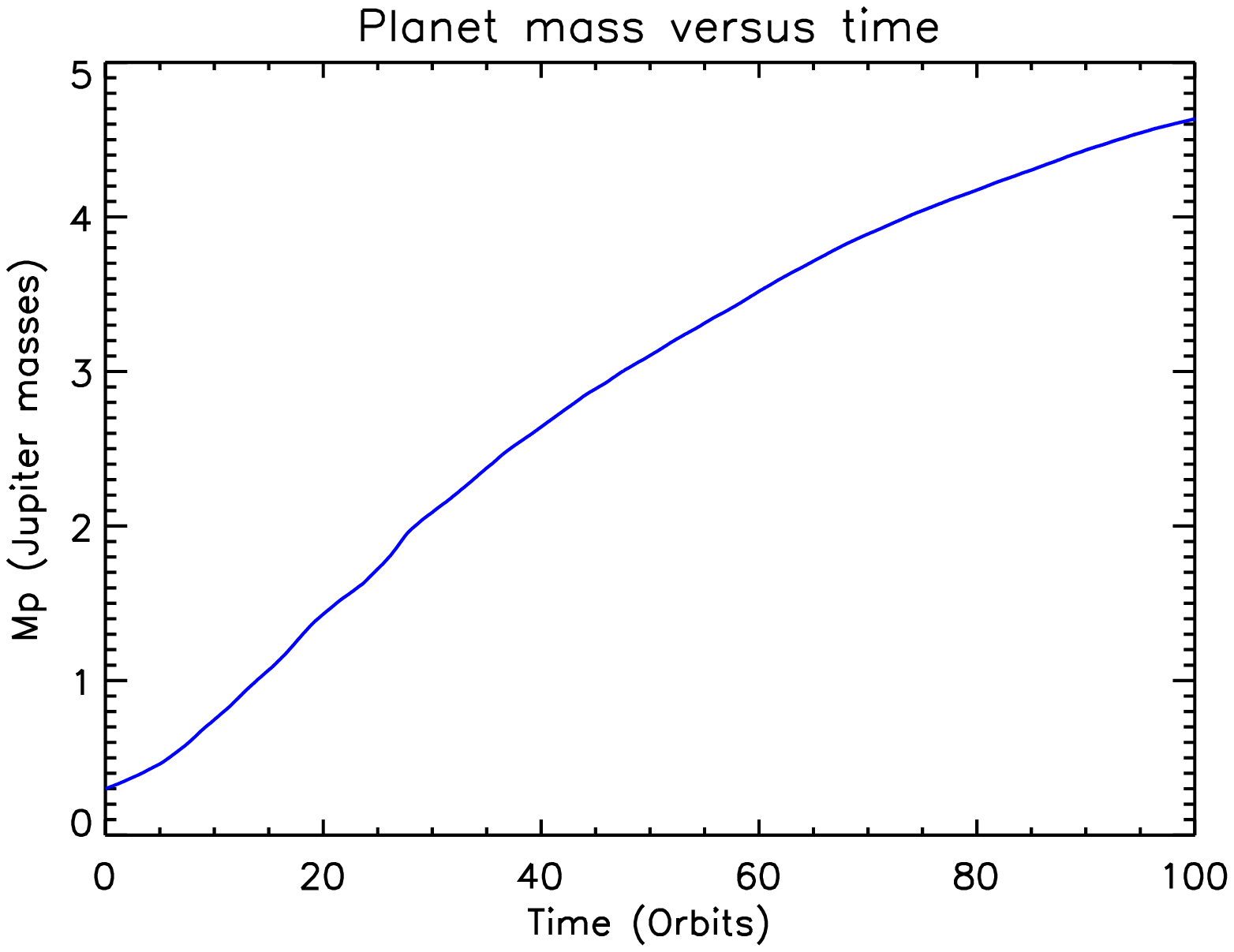}
\caption{\label{turb-fig7} The left panel shows the torque experienced by
the giant as a function of time. The right panel shows the evolution
of the giant planet mass as it accretes gas from the disc.}
\end{figure*}

A number of interesting features arise in simulations of gap forming
planets embedded in turbulent discs, which differentiate
them from runs performed using laminar but viscous disc
models. Comparing figure~\ref{turb-fig5} with figure~\ref{Giant-gap}
shows that the spiral arms in the turbulent disc appear less well-defined
and more diffused than in the laminar models. This point is reinforced
by the left panel of figure~\ref{turb-fig7}, which contains a
snapshot of the planet-disc system in the $r$--$\phi$ plane.
The magnetic field topology is significantly modified in the vicinity
of the protoplanet, as illustrated by the right panel in
figure~\ref{turb-fig7} which displays a snapshot of the quantity
$B^2$ in the disc in the vicinity of the planet.
It can be seen that the magnetic field is compressed and ordered 
in the postshock region associated
with the spiral wakes, increasing the magnetic stresses there.
Accretion of gas into the protoplanet Hill sphere causes advection of
field into the circumplanetary disc that forms there, creating a strong
magnetic field in the circumplanetary disc, and this field
links between the protoplanetary disc and the circumplanetary disc,
producing an effect of magnetic braking of the circumplanetary material.
Indeed, a direct comparison between a simulation of a 3 M$_{\rm Jup}$
protoplanet in a viscous laminar disc and an equivalent turbulent disc
simulation presented by \cite{nelson04}
suggests that mass accretion onto the planet in a 
magnetised disc may be enhanced
by this effect. Simulations are underway at present to explore this
effect within the 3D stratified model presented in this section.
It is worth noting, however, that these global simulations are of 
modest resolution, especially when considering complex physical
processes occuring within the planet Hill sphere,
and more high resolution work needs to be done to examine these
issues in greater detail.

\section{Summary}
\label{summary}
We have reviewed recent progress in the field of disc-planet 
interactions in the context of orbital migration.
This progress has come about to a large degree from large 
scale two and three dimensional 
simulations that have utilised the most
up to date supercomputer resources. These have allowed 
very high resolution simulations to be performed which
examine the detailed behaviour of the disc in the corotation
region, leading to a fuller understanding of the potential 
influence of the corotation torque in modifying the migration
of low mass and fully embedded planets. 
Simulations have also enabled the study of migration in
non isothermal discs in which radiation transport is included,
showing the importance of relaxing the widely used assumption
that the disc gas can be treated as being locally isothermal.
The inclusion of MHD effects in full 3D global simulations
is now possible, allowing the influence of magnetohydrodynamic
turbulence on planet formation and migration to be studied.

Key questions for the future relate to how these various physical effects
combine to change our current understanding of planetary migration.
For example, the long-term action of corotation torques requires
viscous diffusion to operate in the disc in order to prevent
saturation. An important question is how these corotation
torques operate in discs where the effective
viscous stresses are provided by magnetohydrodynamic turbulence ?
Can the corotation torques remain unsaturated within a dead zone ?
These and other important questions are the subject of on-going 
research. The increasing power of available computing facilities
will continue to facilitate the development of increasingly 
realistic disc models, which will in turn lead to an increase in our
understanding of planetary formation and migration.


%


\bibliographystyle{spmpsci}      
\bibliography{disc-planet.bib}   

%
%

\end{document}